\documentclass[12pt]{article}

\usepackage{draft} 
\usepackage[weather]{ifsym}

\usepackage{cite}
\usepackage{mciteplus}
\usepackage{bbm}

\usepackage{amsmath}
\usepackage{amsfonts}
\usepackage{amsmath}
\usepackage{graphicx}
\usepackage{float}
\usepackage{amssymb}
\usepackage{hyperref}
\usepackage{mathtools}
\usepackage{cleveref}
\usepackage{extarrows}
\usepackage{relsize}
\usepackage{bbold}
\usepackage{xcolor}
\newcommand{\pd}[1]{\partial}
\newcommand{\tx}[1]{\tilde{x}}
\newcommand{\beq}{\begin{equation}}
\newcommand{\eeq}{\end{equation}}

\newcommand{\tT}{\mathcal{T}}

\newcommand{\cH}{\mathcal{H}}
\newcommand{\cT}{\mathcal{T}}

\newcommand{\Vc}{\mathcal{C}_4}
\newcommand{\eps}{\varepsilon}

\newcommand{\rel}{\text{Re}}
\newcommand{\ima}{\text{Im}}

\def\ie{\begin{equation}\begin{aligned}}
\def\fe{\end{aligned}\end{equation}}

\def\Xint#1{\mathchoice
	{\XXint\displaystyle\textstyle{#1}}%
	{\XXint\textstyle\scriptstyle{#1}}%
	{\XXint\scriptstyle\scriptscriptstyle{#1}}%
	{\XXint\scriptscriptstyle\scriptscriptstyle{#1}}%
	\!\int}
\def\XXint#1#2#3{{\setbox0=\hbox{$#1{#2#3}{\int}$}
		\vcenter{\hbox{$#2#3$}}\kern-.5\wd0}}
\def\ddashint{\Xint=}
\def\dashint{\Xint-}

\begin{document}

\unitlength = .8mm

\begin{titlepage}

\begin{center}

\hfill \\
\hfill \\
\vskip 1cm

\title{Anyon Scattering from Lightcone Hamiltonian: \\
the Singlet Channel}

\author{Barak Gabai${}^{\text{\Snow}}$, Joshua Sandor${}^{\text{\Snow}}{}^{\text{\Sun}}$, Xi Yin${}^{\text{\Snow}}$}


\address{${}^{\text{\Snow}}$Jefferson Physical Laboratory, Harvard University, \\
Cambridge, MA 02138 USA
\\
${}^{\text{\Sun}}$Stanford Institute for Theoretical Physics, Stanford University, 
\\
Stanford, CA 94305, USA}

\email{bgabai@g.harvard.edu, jsandor@fas.harvard.edu, xiyin@fas.harvard.edu}
\end{center}

\abstract{We study $U(N)$ Chern-Simons theory coupled to massive fundamental fermions in the lightcone Hamiltonian formalism. Focusing on the planar limit, we introduce a consistent regularization scheme, identify the counter terms needed to restore relativistic invariance, and formulate scattering theory in terms of unambiguously defined asymptotic states. We determine the $2\to 2$ planar S-matrix element in the singlet channel by solving the Lippmann-Schwinger equation to all orders, establishing a result previously conjectured in the literature. 
}

\vfill

\end{titlepage}

\eject

\begingroup
\hypersetup{linkcolor=black}

\tableofcontents

\endgroup

\section{Introduction}

Relativistic non-Abelian anyons described by Chern-Simons-matter (CSM) theories in 2+1 dimensions are intriguing for a variety of reasons \cite{Deser:1981wh, Deser:1982vy, Witten:1988hf, Fredenhagen:1988fj, Frohlich:1990ww, Frohlich:1990xz, Frohlich:1991wb, Iengo:1991zbc, Kitaev:2005hzj, Aharony:2011jz, Giombi:2011kc, Aharony:2012ns, Jain:2013gza, Jain:2014nza, Aharony:2015mjs}. Despite the presence of long range interaction, there appear to be well-defined asymptotic states, whose S-matrix elements are subject to unusual analyticity and crossing relations \cite{Jain:2014nza}. To understand the latter would be particularly important for an S-matrix bootstrap approach to questions concerning strongly coupled anyons \cite{Kruczenski:2022lot}.

In this paper, we consider $U(N)$ Chern-Simons theory coupled to a massive fermion field in the fundamental representation, described by the action\footnote{Details of the convention will be given in section \ref{sec:classical}.}
\ie\label{CSAction}
S = \int d^3x \Big[-\frac{k}{4\pi}\,\tr\Big(A\wedge dA + \frac{2i}{3} A\wedge A\wedge A\Big)- i \overline{\psi} \gamma^\mu D_\mu \psi - i m_0 \overline{\psi}\psi\Big].
\fe
The conventional Feynman diagram approach to scattering theory \cite{Jain:2014nza} has its origin in LSZ reduction, through which the S-matrix elements are extracted from Green functions. However, the gauge-invariant operators that create anyons are necessary non-local, e.g. the fermion field $\psi$ attached to a Wilson line of the Chern-Simons gauge field. One might suspect subtleties in the definition of asymptotic states, or even the existence of the LSZ limit of the appropriate correlation functions with Wilson lines, due to the long-range nature of the Chern-Simons gauge interaction.

The first objective of this paper is to formulate the asymptotic states, and their scattering amplitudes, in an unambiguous manner. This is achieved in the lightcone Hamiltonian formalism, based on quantization of (\ref{CSAction}) in the lightcone gauge. We introduce a regularization scheme that involves a UV cutoff on the fermion momentum transverse to the lightcone, an IR cutoff on the fermion lightcone momentum, and a principal value prescription that regularizes the Chern-Simons propagator at zero lightcone momentum. In such a scheme, we will identify the counter terms that restores the underlying 2+1 dimensional Poincar\'e symmetry, in the planar limit. The asymptotic states and scattering theory can then be formulated using Lippmann-Schwinger equations.

The $2\to 2$ S-matrix element of particle-anti-particle scattering is expected to take the general form
\ie\label{sanygen}
& S_{i \bar{j}}^{k \bar{\ell}}(\vec{p}_3,\vec{p}_4 | \vec{p}_1,\vec{p}_2) = {}^{out}\langle k, \vec p_3;\bar \ell,\vec p_4 |i,\vec p_1; \bar j, \vec p_2 \rangle^{in}
\\
&= \delta _i^k\delta^{\bar{\ell}}_{\bar{j}} I(p_3,p_4| p_1,p_2) 
- i   (2\pi)^3 \delta^3(p_1+p_2-p_3-p_4) \bigg[ \Big(\delta _i^k\delta^{\bar{\ell}}_{\bar{j}}-\frac{1}{N}\delta_{i\bar j}\delta^{k\bar \ell} \Big) T_{\text{A}}(s,\theta) \\&\quad\quad\quad\quad\quad\quad\quad\quad\quad\quad\quad\quad\quad\quad\quad\quad\quad\quad\quad\quad\quad\quad\quad\quad\quad\quad\quad\quad+ \frac{1}{N} \delta_{i\bar j}\delta^{k\bar \ell} T_\text{S}(s,\theta) \bigg],
\fe
where $i,j$ and $\bar k, \bar \ell$ are gauge indices for fermions in the fundamental and anti-fundamental representations. $p_a=(E_a,\vec p_a)$ is the energy-momentum of the $a$-th particle. $I(p_3,p_4|p_1,p_2)$ is the identity matrix element
\ie\label{identityi}
I(p_3,p_4| p_1,p_2)  &= (2\pi)^4 (2E_1)(2 E_2) \delta^2(\vec p_{13}) \delta^2(\vec p_{24}) 
\\
&= (2\pi)^3 \delta^3(p_1+p_2-p_3-p_4)\cdot 8\pi\sqrt{s} \delta(\theta),
\fe
where $s\equiv -(p_1+p_2)^2$, and $\theta$ is the scattering angle related by  $t\equiv -p_{13}^2=-(s-4m^2){1-\cos\theta\over 2}$.
$T_{\rm A}(s,\theta)$ and $T_{\rm S}(s,\theta)$ are the connected amplitudes in the gauge adjoint and singlet channel respectively.

An unusual feature of the anyon S-matrix is its non-standard cluster property. In particular, the connected amplitude in (\ref{sanygen}) contains a distribution supported in the forward direction $\theta=0$. It is only after subtracting off the latter that one can speak of the analyticity and crossing properties of the amplitude. Focusing on the singlet channel in the planar limit, it was conjectured in \cite{Jain:2014nza} that $T_{\rm S}$ takes the form
\ie\label{ttapp}
T_{\rm S}(s,\theta) = T_0(s) \delta(\theta) + \widetilde T(s,\theta),
\fe
where\footnote{Our convention is related to that of \cite{Jain:2014nza} by $m=-c_f$ and a different overall sign for $T$-matrix (\ref{sanygen}) which is natural from the Lippmann-Schwinger relation.}
\ie\label{ttconjres}
& T_0(s) =  8\pi i\sqrt{s} \big(\cos(\pi\widetilde\lambda)-1\big) ,
\\
& \widetilde T(s,\theta) =-  4 i \sqrt{s} \sin\big(\pi  \widetilde{\lambda }\big) \left[ \cot {\theta\over 2} -i \frac{ 1 + e^{i \pi \widetilde{\lambda } } \left({\sqrt{s}+2 m\over\sqrt{s}-2 m}\right){}^{\widetilde{\lambda }+1}}{1-e^{i \pi \widetilde{\lambda }} \left(\frac{\sqrt{s}+2 m}{\sqrt{s}-2 m}\right){}^{\widetilde{\lambda }+1}}\right].
\fe
Here $\widetilde\lambda=N/k_{\rm dr}$ is the 't Hooft coupling, with the Chern-Simons level $k_{\rm dr}$ defined in the dimensional reduction scheme.\footnote{$k_{\rm dr}$ is related to the Chern-Simons level $\kappa$ in the Yang-Mills regularization scheme by $k_{\rm dr} = \kappa + N \,{\rm sign}(\kappa)$ \cite{Giombi:2011kc, Jain:2014nza}.}

In the lightcone Hamiltonian formulation of the scattering problem, the 2+1 dimensional Lorentz symmetry is not manifest. It suffices to work in a sector of fixed $s$, the total lightcone momentum $P^+$, and zero total momentum transverse to the lightcone directions. We can label the asymptotic particles by their momenta in the lightcone and transverse directions, $\overline p_i \equiv (p_i^+,p_i^\perp)$, and pass to the kinetic variables to $(x, y, p, q)$, related to $\overline p_i$ by
\ie\label{momentasplit}
& \overline p_1 = ((1-x)P^+, -p),~~~~ \overline p_2 = (x P^+, p),~~~~ \overline p_3 = ((1-y) P^+, -q),~~~~ \overline p_4 = (y P^+, q),
\fe
where $x,y\in (0,1)$ parameterize the fraction of lightcone momenta distributed between the two particles in the in- and out-state respectively. Note that $P^+$ rescales under Lorentz boost in the lightcone direction, and will drop out of Lorentz invariant observables. $p$ and $q$, on the other hand, are determined by $x, y, P^+$, and $s$ up to a pair of signs, namely
\ie\label{pqsrel}
p = \pm \sqrt{s x(1-x) - m^2},~~~~ q = \pm \sqrt{s y(1-y) - m^2}.
\fe

\begin{figure}[h!]
	\def\svgwidth{0.8\linewidth}
	\centering{
\begingroup%
  \makeatletter%
  \providecommand\color[2][]{%
    \errmessage{(Inkscape) Color is used for the text in Inkscape, but the package 'color.sty' is not loaded}%
    \renewcommand\color[2][]{}%
  }%
  \providecommand\transparent[1]{%
    \errmessage{(Inkscape) Transparency is used (non-zero) for the text in Inkscape, but the package 'transparent.sty' is not loaded}%
    \renewcommand\transparent[1]{}%
  }%
  \providecommand\rotatebox[2]{#2}%
  \newcommand*\fsize{\dimexpr\f@size pt\relax}%
  \newcommand*\lineheight[1]{\fontsize{\fsize}{#1\fsize}\selectfont}%
  \ifx\svgwidth\undefined%
    \setlength{\unitlength}{629.29133858bp}%
    \ifx\svgscale\undefined%
      \relax%
    \else%
      \setlength{\unitlength}{\unitlength * \real{\svgscale}}%
    \fi%
  \else%
    \setlength{\unitlength}{\svgwidth}%
  \fi%
  \global\let\svgwidth\undefined%
  \global\let\svgscale\undefined%
  \makeatother%
  \begin{picture}(1,0.27477477)%
    \lineheight{1}%
    \setlength\tabcolsep{0pt}%
    \put(0,0){\includegraphics[width=\unitlength,page=1]{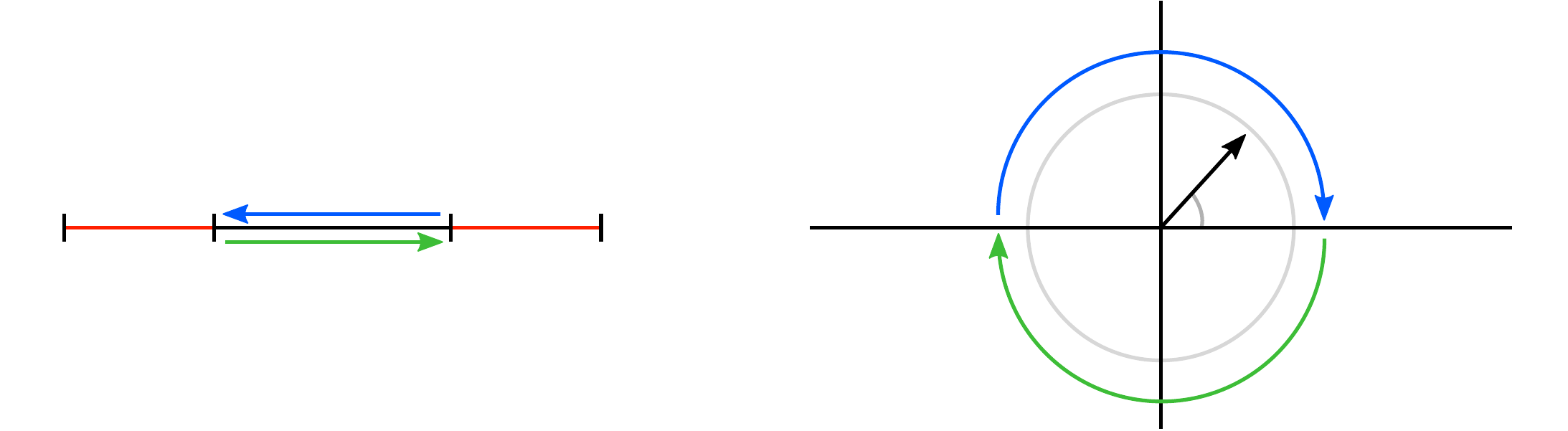}}%
    \put(0.00633121,0.08768836){\makebox(0,0)[lt]{\lineheight{1.25}\smash{\begin{tabular}[t]{l}$x=0$\end{tabular}}}}%
    \put(0.34876787,0.08768833){\makebox(0,0)[lt]{\lineheight{1.25}\smash{\begin{tabular}[t]{l}$x=1$\end{tabular}}}}%
    \put(0.27880516,0.15614328){\makebox(0,0)[lt]{\lineheight{1.25}\smash{\begin{tabular}[t]{l}$x_*^+$\end{tabular}}}}%
    \put(0.11749505,0.15614278){\makebox(0,0)[lt]{\lineheight{1.25}\smash{\begin{tabular}[t]{l}$x_*^-$\end{tabular}}}}%
    \put(0.93427586,0.0924556){\makebox(0,0)[lt]{\lineheight{1.25}\smash{\begin{tabular}[t]{l}$p_1^1$\end{tabular}}}}%
    \put(0.75066451,0.25923935){\makebox(0,0)[lt]{\lineheight{1.25}\smash{\begin{tabular}[t]{l}$p_1^2$\end{tabular}}}}%
    \put(0.77397211,0.13759262){\makebox(0,0)[lt]{\lineheight{1.25}\smash{\begin{tabular}[t]{l}$\alpha$\end{tabular}}}}%
  \end{picture}%
\endgroup%
	\caption{Path of the angle $\alpha$ as $x$ varies in the physical region $[x_*^-, x_*^+]$. The red ``unphysical" region on the interval plays a role in the integral equation (\ref{shortbestintegraleqn}) or (\ref{bestintegraleqn}).
			\label{anglechanging}
	}}
\end{figure}
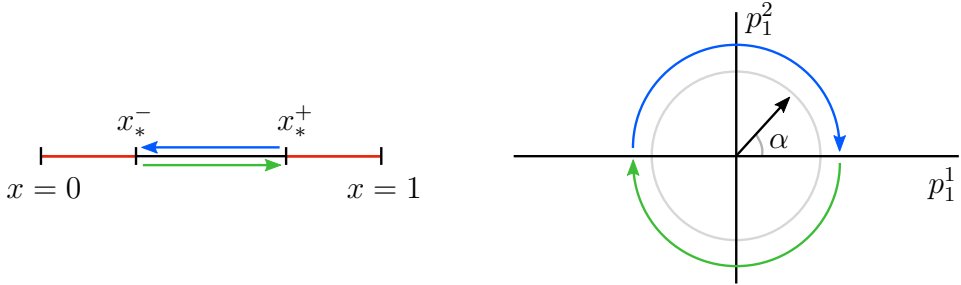

Working at fixed $s$, we can label the in-state with $(x,a)$, where $a\equiv {\rm sign}(p)$($=\pm$), and likewise label the out-state with $(y,b)$, $b\equiv {\rm sign}(q)$. The singlet channel amplitude can be written as a distribution of the form 
\ie\label{eq:genform}
{\cal T}^{b a}(y|x) = \delta(x-y) \delta^{ba} {\cal T}_0(x) + \widetilde {\cal T}^{ba}(y|x).
\fe
Note that $x$ and $y$ take value in the range $[x_*^-, x_*^+]$, where $x_*^\pm = {1\over 2} \pm {1\over 2} \sqrt{s-4m^2\over s}$. The relation between $x$ and the incoming angle $\A$ of the particle of momentum $p_1$ is portrayed in Figure \ref{anglechanging}. The relation between $y$ and the outgoing angle of the particle of momentum $p_3$ is similar.

In the planar limit, the Lippman-Schwinger equation in the singlet sector involves only two-particle intermediate states. We will be able to reduce it to an integral equation over a single variable, of the form 
\ie\label{shortbestintegraleqn}
{\cal T}^{b a}(y|x) = {\cal V}^{ba}(y|x) + \dashint_0^1 dz\sum_{c=\pm} {\cal K}^{bc}(y|z) {\cal T}^{ca}(z|x).
\fe
Importantly, the distribution $\cT^{ba}(y|x)$ appearing on both sides of this equation is an extension of the physical amplitude (\ref{eq:genform}) to the domain $y\in (0,1)$, which includes unphysical kinematic region corresponding to complex angles of out-particles. Details of the function ${\cal V}^{ba}(y|x)$ and the kernel ${\cal K}^{bc}(y|z)$ will be described in section \ref{sec:integraleqn}. 
After a careful inspection of the forward scattering singularity, we will see that the solution to (\ref{shortbestintegraleqn}) is related to (\ref{ttapp}) and (\ref{ttconjres}) by
\ie\label{ttmatch}
{\cal T}^{ba}(y|x) =  ({\cal N}(\B))^*\, {\cal N}(\A) \, T_{\rm S}(s,\theta=\beta-\alpha),
\fe
provided a finite coupling renormalization between the 't Hooft coupling $\lambda$ appearing in the lightcone Hamiltonian, and $\widetilde\lambda$ appearing in (\ref{ttconjres}),
\ie\label{couplingredef}
\lambda = {2(1-\cos(\pi\widetilde\lambda))\over \pi \sin(\pi\widetilde\lambda)}.
\fe
$\alpha$ and $\beta$ are the angles of the incoming particle momentum $p_1$ and the outgoing particle momentum $p_3$ with respect to the lightcone directions, defined by (\ref{comassignment}) or equivalently (\ref{eq:xtoangle}).
The prefactor  $({\cal N}(\B))^*\, {\cal N}(\A)$ appearing on the RHS of (\ref{ttmatch}) is due to a different normalization and phase convention for the asymptotic states in the Lippmann-Schwinger formalism from those of the covariant amplitude. 
The result (\ref{ttmatch}) thus confirms the conjectured the singlet channel planar S-matrix of \cite{Jain:2014nza}. 

In section \ref{sec:classical}, we derive the lightcone Hamiltonian of the CSM theory in the lightcone gauge. Details of regularization scheme and counter terms in the quantum Hamiltonian are discussed in section \ref{sec:quantumH}. In section \ref{sec:integraleqn}, we formulate the scattering theory using Lippmann-Schwinger equation, restricted to the 2-particle gauge singlet sector in the planar limit. In section \ref{sec:solution}, starting with the 1-loop approximation, and then analyzing the forward limit singularity, we will be able to identify the solution to L-S equation to all orders in $\lambda$. We conclude with future perspectives in section \ref{sec:discuss}. Further technical details including a numerical verification of the solution to the integral equation are given in the Appendices.

\section{The lightcone Hamiltonian in the lightcone gauge}
\label{sec:classical}

We will formulate the 2+1 dimensional $U(N)$ Chern-Simons theory coupled to a Dirac fermion field in the fundamental representation through the lightcone Hamiltonian in close parallel to that of the 1+1 dimensional 't Hooft model \cite{tHooft:1974pnl, Dalley:1992yy, Bhanot:1993xp,Dempsey:2021xpf,Giombi:2011kc}.

\subsection{The classical lightcone Hamiltonian}

Our convention for lightcone coordinates is $x^{\pm} = \frac{1}{\sqrt{2}}(\pm x^0+ x^1)$, and the transverse coordinate will be denoted $x^\perp = x^2$. The classical action is given by (\ref{CSAction}), where the gauge covariant derivative is defined as $D_\mu = \partial_\mu - i A^a_\mu t^a$. In the lightcone gauge
\ie
A_-=0,
\fe
the Faddeev-Popov ghosts decouple, and the action can be written as\footnote{We adopt the gamma matrix convention \ie
\gamma^+ = \begin{pmatrix}
	0& \sqrt{2} \\
	0 & 0
\end{pmatrix}\,,\quad
\gamma^- = \begin{pmatrix}
	0& 0\\
	\sqrt{2}  & 0
\end{pmatrix}\,,\quad
\gamma^\perp = \begin{pmatrix}
	1& 0 \\
	0 & -1
\end{pmatrix}\,,\quad
\gamma^0 = \begin{pmatrix}
	0& 1\\
	-1 & 0
\end{pmatrix}\,.
\fe} 
\begin{multline}\label{CSMAction}
\cS = \int dx^+ dx^- dx^3 \Big[-\frac{k}{8\pi}\epsilon^{ij}A^a_i \partial_- A_j^a + i \psi_-^\dagger D_+ \psi _- -i  \psi_+^\dagger D_- \psi_+ \\+ \frac{i}{\sqrt{2}}(\psi_+^\dagger D_\perp \psi_- + \psi_-^\dagger D_\perp \psi_+)-\frac{i}{\sqrt{2}}  m_0 (\psi_+^\dagger \psi_--\psi_-^\dagger \psi_+ )\Big] \,,
\end{multline}
where we have used $\overline{\psi} =  \psi^\dagger \gamma^0$, and have redefined $\psi =2^{-1/4}(\psi_+,\psi_-)$. $\epsilon^{ij}$ is the constant anti-symmetric tensor with $\epsilon^{\perp+}=1$.

To proceed, we will view $x^+$ as the time coordinate. The absence of kinetic terms for the gauge fields $A_+$ and $A_\perp$ means that the latter are non-dynamical fields, and can be eliminated by solving their equations of motion
\begin{align}
\partial_- A^a_\perp &= -\frac{4\pi}{k}\psi_{-}^\dagger t^a\psi_{-}\,,\nonumber\\
\partial_- A^a_+ &= \frac{4 \pi}{\sqrt{2}k}\Big( \psi_-^\dagger t^a \psi_++\psi_+^\dagger t^a\psi_-   \Big)\,.
\end{align} 
In addition, the fermion field components $\psi_+$, $\psi_+^\dagger$ are also non-dynamical, and be eliminated through their equations of motion
\begin{align}\label{psiplus}
i\sqrt{2}\partial_- \psi_{+,j} &= i\partial_\perp \psi_{-,j} - \frac{2\pi }{k} \Big( \frac{1}{\partial_-}\psi_{-,\bar\ell}^\dagger\psi_{-,j}\Big)\psi_{-,\ell} -i m_0 \psi_-\,.
\end{align}
This leaves $\psi_-$, $\psi_-^\dagger$ as the only dynamical fields. Here and henceforth we adopt a notation in which a lower anti-fundamental gauge index $\bar\ell$ is equivalent to an upper fundamental index $\ell$.

The path integral can then be put in Hamiltonian form, with the classical lightcone Hamiltonian given by\footnote{Here we used the $U(N)$ completeness relation $(t^a)_k{}^\ell (t^a)_j{}^i = \frac{1}{2} \delta_k^i\delta_j^\ell $. For $SU(N)$ gauge group, additional $\mathcal{O}(N^{-1})$ terms would appear.} 
\ie\label{lightconeMomentum}
H &= - P^- =  \int dx^-dx^\perp\, T_{-+}  \nonumber\\ 
&= - \int dx^- dx^\perp\, \frac{1}{\sqrt{2}}\Bigg[ i \psi^\dagger_- \partial_\perp \psi_+ + im_0\psi_-^\dagger \psi_+ - \frac{2\pi }{k}\,\Big(\frac{1}{\partial_-}\psi_{-,\bar j}^\dagger \psi_{-,i} \Big)(\psi_{-,\bar i}^\dagger\psi_{+,j}  )\Bigg]\,,
\fe
where in the second line $\psi_+$ is understood to be replaced by the solution to (\ref{psiplus}). 
It is convenient to work with the Fourier transformed fields
\ie\label{psifour}
\psi_{-,i}(x) = \int \frac{d^2\overline p}{(2\pi)^2} e^{i(p^+ x^- + p^\perp x^\perp)} \widetilde\psi_{-,i}(x^+,\overline p), ~~~~ \psi_{-,\bar i}^\dagger(x)  =  \int \frac{d^2\overline p}{(2\pi)^2} e^{-i(p^+ x^- + p^\perp x^\perp)} \widetilde\psi_{-,\bar i}^\dagger(x^+,\overline p) ,
\fe
where $\overline p \equiv (p^+,p^\perp)$. In writing expressions of the Hamiltonian below, we will always be working at a fixed time $x^+$, and omit the explicit dependence on $x^+$ in the fields. 
The lightcone Hamiltonian can be decomposed as 
\ie\label{h246}
H= H_2 + H_4 + H_6,
\fe
where
\ie
\label{eq:Hclassical}
H_2 &= \int \frac{d^2\overline p}{(2\pi)^2}\,h_2(\overline p;m_0)\, \left[ \psi_-^\dagger(\overline p)\psi_-(\overline p)\right],
\\
H_4&= \int \prod_{i=1}^4 \frac{d^2\overline p_i}{(2\pi)^2}\,(2\pi)^2 \delta^2(\overline p_{12}+\overline p_{34}) \,h_4(\overline p_1,\overline p_4;m_0)\left[ \psi_-^\dagger(\overline p_1)  \psi_- (\overline p_2) \right]\left[ \psi_-^\dagger(\overline p_3)\psi_-(\overline p_4) \right] ,
\\
H_6 &= \int \prod_{i=1}^6 \frac{d^2\overline p_i}{(2\pi)^2}\,(2\pi)^2 \delta^2\Big(\overline p_{12}+\overline p_{34}+\overline p_{56}\Big)\, h_6(\overline p_2,\overline p_3,\overline p_4,\overline p_5,\overline p_6)
\\
&~~~~~~~~~~~~~\times  \left[ \psi_-^\dagger(\overline p_1)\psi_-(\overline p_2) \right] \left[ \psi_-^\dagger(\overline p_3)\psi_-(\overline p_4) \right] \left[ \psi_-^\dagger(\overline p_5)\psi_-(\overline p_6) \right] .
\fe
Here the color indices are contracted between fermion fields in the bracket, i.e. $[\psi^\dagger \psi] \equiv \psi^\dagger_{\bar i} \psi_i$. The coefficients $h_2, h_4, h_6$ are given by \cite{Giombi:2011kc, Delacretaz:2018xbn}
\ie\label{Hcoeff}
h_2(\overline p;m_0) &= \frac{(p^\perp)^2+m_0^2}{2p^+},
\\
h_4(\overline p_1,\overline p_4;m_0)&= {\pi\over k} {1\over (p_{14}^+)_\eps} \Bigg( \frac{ip_1^\perp + m_0}{ p_1^+}+\frac{ip_4^\perp -m_0}{ p_4^+} \Bigg),
\\
h_6(\overline p_2,\overline p_3,\overline p_4,\overline p_5,\overline p_6) &= -\frac{2\pi^2}{k^2}\frac{1}{(p_{23}^+)_\eps(p_{45}^+)_\eps(p_{45}^++p_6^+)_\eps}.
\fe
where we have introduce the notation
\ie\label{smoothcutoff}
\frac{1}{(p^+)_\varepsilon} \equiv \frac{p^+}{(p^+)^2 + \varepsilon^2}
\fe 
for IR-regulated propagators. The latter will lead to a principal value prescription for lightcone momentum integrals appearing in scattering amplitudes.

\subsection{Regularization scheme and counter terms}
\label{sec:quantumH}

The time evolution defined by the path integral based on the classical Hamiltonian $H$ given in (\ref{lightconeMomentum}) or (\ref{h246}) is equivalent to that of a quantum lightcone Hamiltonian operator $\widehat\cH$. Modulo potential operator ordering and regularization ambiguities, $\widehat\cH$ is related to $H$ by promoting $\psi_-, \psi_-^\dagger$ to field operators subject to the (equal $x^+$-time) canonical quantization relation
\ie
\Big\{(\psi_-)_{\bar i}^\dagger(x^-, x^\perp),(\psi_-)_j(y^-,y^\perp)\Big\} = \delta_{\bar ij}\delta^2(x- y)\,,
\fe 
We will separate the Fourier transformed field operators $\widetilde\psi_{-,i}(x^+,\overline  p)$, $\widetilde\psi_{-,\bar i}^\dagger(x^+,\overline p)$, related by (\ref{psifour}), into positive and negative frequency modes according to
\ie\label{eq:creanni}
& \widetilde\psi_{-,i}(\overline p) = \Theta (p^+) a_i(\overline p) + \Theta(-p^+) b^\dagger_i(-\overline p),
~~~~\widetilde \psi_{-,\bar i}^\dagger(\overline p) = \Theta (p^+) a_{\bar i}^\dagger(\overline p) + \Theta(-p^+) b_{\bar i}(-\overline p),
\fe
where $a_i(\overline p)$ and $b_{\bar i}(\overline p)$ can be viewed as fermion annihilation operators (defined for $p^+>0$ and arbitrary $p^\perp$) that obey
\ie\label{eq:1pnorm}
\Big\{a^\dagger_{\bar i}(\overline p), a_{j}(\overline q)\Big\} = \Big\{b^\dagger_j(\overline p), b_{\bar i}(\overline q)\Big\} = \delta_{\bar ij} (2\pi)^2 \delta^2(\overline p-\overline q) .
\fe
To write the precise expression of $\widehat\cH$ in terms of the fermion creation and annihilation operators requires a choice of regularization scheme. We will adopt a scheme in which the fermion modes are subject to both IR and UV cutoff on their lightcone momentum $p^+$, and a UV cutoff on their transverse momentum $p^\perp$, according to 
\ie\label{ppscheme}
\delta< p^+< \widetilde\delta^{-1},~~~~~ |p^\perp|<\Lambda,
\fe
and will eventually take the limit $\delta,\widetilde\delta\to 0^+$, $\Lambda\to \infty$ in determining physical observables.
The naive replacement of fermion fields in $H$ by their corresponding field operators, with a given choice of ordering, promotes $H$ to a quantum operator $\widehat H_{\rm naive}$. The true quantum Hamiltonian $\widehat\cH$ may in principle differ from $\widehat H_{\rm naive}$ by counter terms that take the form of operator ordering ambiguities. Due to the renormalizability of CSM theory, such counter terms are in principle fixed by the requirement Lorentz invariance and locality. 

More explicitly, we can expand
\ie\label{lchamil}
\widehat{\cal H} = \widehat{\cal H}_2 + \widehat{\cal H}_4+ \widehat{\cal H}_6,
\fe
where each $\widehat {\cal H}_n$ is a linear combination of normal-ordered products of $n$ fermion creation operators $a^\dagger, b^\dagger$ and/or annihilation operators $a,b$. In particular, the ``free-particle" part of the quantum lightcone Hamiltonian, $\widehat\cH_2$, takes the form
\ie
\widehat{\cal H}_2 &=    \int \frac{d^2\overline p}{(2\pi)^2} \Theta(p^+) \,\cH_2(\overline p)\Big[ a_{\bar \ell}^\dagger(\overline p) a_\ell(\overline p) + b_\ell^\dagger(\overline p) b_{\bar \ell}(\overline p)\Big] ,
\fe
where
\ie\label{eq:H2}
\cH_2(\overline p) =h_2(\overline p;m)= \frac{(p^\perp)^2+m^2}{2 p^+}.
\fe
Here $m$ is the physical ``renormalized" mass of the fermion. 

$\widehat\cH_4$ can be decomposed as
\ie
\widehat{\cal H}_4 = \widehat{\cal H}^\text{S}_4+\widehat{\cal H}_4^\text{A}+\widehat{\cal H}_4^\text{P-P} +\widehat{\cal H}_4^\text{A-A} +\widehat{\cal H}_4^{3\to 1}+\widehat{\cal H}_4^{1\to 3} ,
\fe
where $\widehat{\cal H}^\text{S}_4$ and $\widehat{\cal H}_4^\text{A}$ represent particle/anti-particle interaction in the gauge singlet and adjoint channels respectively, $\widehat{\cal H}_4^\text{P-P}$ and $\widehat{\cal H}_4^\text{A-A}$ represent particle/particle interaction and anti-particle/anti-particle interaction respectively, $\widehat{\cal H}_4^{3\to 1}$ and $\widehat{\cal H}_4^{1\to 3}$ represent interactions that change particle number by $\mp 2$. Further details are described in appendix \ref{app:Ham}. 

$\widehat\cH_6$, on the other hand, is free of ordering ambiguities, and is given by $H_6$ of (\ref{eq:Hclassical}) with all fields promoted to quantum operators and normal ordered.

For the rest of this paper, we will restrict to the planar limit, defined as $N, k\to \infty$, with the 't Hooft coupling $\lambda=N/k$ fixed. In this limit, particle production is suppressed. Furthermore, in the 2-particle singlet sector, the only part of the lightcone Hamiltonian that affects the planar S-matrix element is
\ie\label{h24sing}
\widehat\cH_2 + \widehat\cH_4^{\rm S}.
\fe

\begin{figure}[h!]
	\def\svgwidth{0.8\linewidth}
	\centering{
		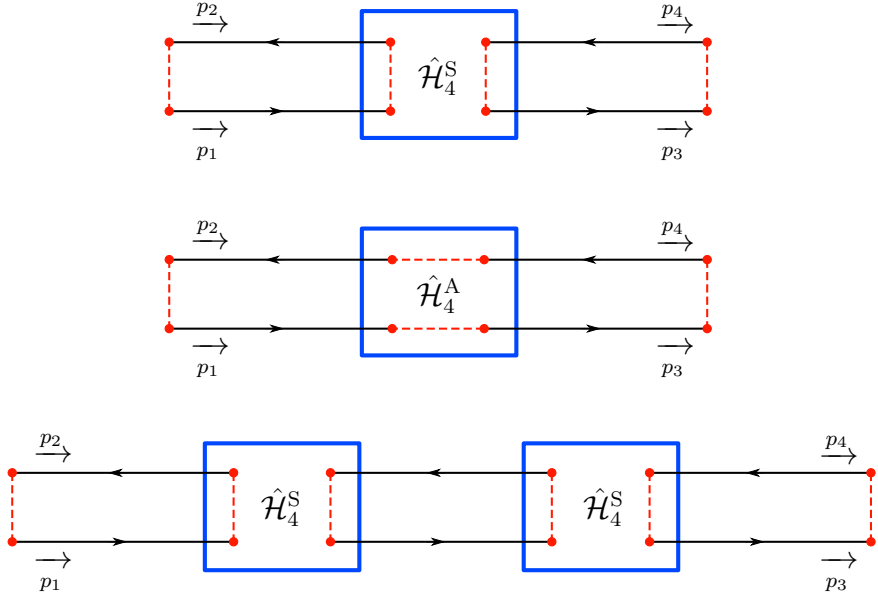
		\caption{Tree-level leading and subleading contributions and one-loop leading contribution to scattering in the Lippmann-Schwinger formulation. 
			\label{treeimg}
	}}
	
\end{figure}

In the scheme defined by the regulators (\ref{smoothcutoff}) and (\ref{ppscheme}), we propose that $\widehat {\cal H}$, at least in the planar limit, is given precisely by $\widehat H_{\rm naive}$ defined with the same ``naive" ordering as seen in the expression (\ref{eq:Hclassical}). In other words, all counter terms contained in $\widehat\cH_2$ and $\widehat\cH_4$ come from the difference between $\widehat H_{\rm naive}$ and its normal-ordered version. This results in the physical mass\footnote{Note that the re-ordering contributions from quartic and sextic terms in the field operators combine to give $m^2$. Furthermore, the dependence on $\widetilde\delta$ is entirely absorbed by a choice of counter term in $m_0$, and will no longer appear in our formulation of scattering theory at fixed total lightcone momentum $P^+$.}
\ie
m = m_0 + 2  \pi \lambda \, \int_{\delta<p^+<\widetilde\delta^{-1}, ~|p^\perp|<\Lambda} \frac{d^2\overline p}{(2\pi)^2}\,\frac{1}{p^+} ,
\fe
and the quartic singlet interaction term
\ie\label{interaction}
\widehat{\cal H}_4^\text{S} &=   \int \prod_{i=1}^4 \frac{d^2\overline p_i}{(2\pi)^2} \Theta(p_i^+) \,(2\pi)^2 \delta^2(\overline p_{14}+\overline p_{23}) 
\\
&~~~~~~~\times\frac{1}{N}\cH_4^{\rm reg}(\overline p_1,\overline p_2,\overline p_3,\overline p_4; \delta,\Lambda)\, a^\dagger_{\bar k}(\overline p_3) b^\dagger_k(\overline p_4)  b_{\bar \ell}(\overline p_2) a_\ell(\overline p_1),
\fe
where the coefficient ${\cal H}_4^{\rm reg}$ given by
\ie
\label{eq:H4}
\cH_4^{\rm reg}(\overline p_1,\overline p_2,\overline p_3,\overline p_4;\delta,\Lambda) = \cH_4^{\rm cl}(\overline p_1,\overline p_2,\overline p_3,\overline p_4) + \cC_4(\overline p_1,\overline p_2,\overline p_3,\overline p_4;\delta,\Lambda).
\fe
In (\ref{eq:H4}), the ``classical" coupling ${\cal H}_4^{\rm cl}$ comes from $H_4$ in (\ref{eq:Hclassical}), 
\ie\label{H4explicit}
\cH_4^{\rm cl}(\overline p_1,\overline p_2,\overline p_3,\overline p_4) &= h_4(\overline p_3, \overline p_1;m)+ h_4(-\overline p_2,-\overline p_4;m) 
\\
&= - \pi\lambda \left[ \frac{m\, p_{13}^+ +i (p_1^\perp p_3^++p_1^+ p_3^\perp)}{p_1^+ p_3^+ (p_{13}^+)_\varepsilon}+\frac{m \, p_{24}^+ +i (p_2^\perp p_4^++p_2^+ p_4^\perp)}{p_2^+ p_4^+ (p_{24}^+)_\varepsilon}\right] ,
\fe
where the ``counter term" ${\cal C}_4$ comes from the re-ordering of $H_6$ in (\ref{eq:Hclassical}),
\ie\label{c4counter}
&\cC_4(\overline p_1,\overline p_2,\overline p_3,\overline p_4;\delta,\Lambda) = -2\pi^2\lambda^2\,(p_1^+ +p_2^+)  
\\
& \times\int_{p^+> \delta,~ |p^\perp|<\Lambda} {d^2\overline p\over (2\pi)^2} \frac{1}{p^+}  \bigg[ \frac{1}{(p^+-p_4^+)_{\varepsilon}(p_2^+-p^+)_{\varepsilon}(p_1^++p_2^+-p^+)_{\varepsilon}}\\&\quad\quad\quad\quad\quad\quad\quad\quad\quad\quad\quad\quad\quad\quad\quad\quad- \frac{1}{(p_1^++p^+)_\varepsilon(p_3^++p^+)_\varepsilon(p_1^++p_2^++p^+)_\varepsilon}\bigg].
\fe
The two terms in the bracket on the RHS of (\ref{c4counter}) can be viewed as due to exchange of gauge bosons between the fermions, and fermion pair annihilation/creation by gauge bosons, respectively. Note that the $\varepsilon$-regulator is not actually necessary for the propagators appearing in the second term, as the latter is non-singular in the integration domain.

While we do not have an a priori derivation of the proposed counter terms, we will see that the resulting 2-particle singlet channel Hamiltonian (\ref{h24sing}) produces $2\to 2$ scattering amplitudes that are consistent with Lorentz invariance in a highly nontrivial manner.

\section{Scattering equation in the 2-particle singlet sector}
\label{sec:integraleqn}

In this section we formulate the S-matrix in the 2-particle singlet sector using the Lippmann-Schwinger equation based on the planar lightcone Hamiltonian (\ref{lchamil}).


\subsection{Lippmann-Schwinger equation}
\label{sec:lseqn}

We begin by separating the lightcone Hamiltonian into its free part $\widehat \cH_2$ (\ref{eq:H2}) and interacting part $\widehat V=\widehat\cH_4 + \widehat\cH_6$,
\ie
\widehat\cH = \widehat\cH_2 + \widehat V,
\fe
and denote by $|\A\rangle^0$ an eigenbasis with respect to $\widehat\cH_2$ (``free-particle basis"), indexed by $\A$ with $\widehat\cH_2|\A\rangle^0 = E_\A |\A\rangle^0$. The corresponding in- and out- scattering states are related by Lippmann-Schwinger equation
\ie\label{inoutrelation}
|\A\rangle^{\rm in/out} = |\A\rangle^0 + {1\over E_\A - \widehat \cH_2 \pm i\epsilon} \widehat V |\A\rangle^{\rm in/out},
\fe
The S-matrix elements can be written as
\ie\label{smatrixeqn}
S(\B|\A)\equiv {}^{\rm out}\langle \B|\A\rangle^{\rm in} = {}^0\langle\B|\A\rangle^0 - 2\pi i \delta(E_\B-E_\A) \widehat {\cal T}(\B|\A),
\fe
where $\widehat {\cal T}(\B|\A)$ is given by (see Appendix  \ref{app:tmatrix} for a derivation of this standard fact)
\ie\label{tmatrixexpr}
\widehat {\cal T}(\B|\A) = {}^0\langle \B| \widehat V|\A\rangle^{\rm in}.
\fe
Note that the $T$-matrix element (\ref{tmatrixexpr}) is defined without imposing energy conservation, and is an extension of the physical amplitude appearing in (\ref{smatrixeqn}). It follows from (\ref{inoutrelation}) that $\widehat {\cal T}(\B|\A)$ obeys the integral equation
\ie\label{hattintegraleqn}
\widehat{\cal T}(\B|\A) = {}^0\langle\B| \widehat V |\A\rangle^0 + \int d\C { {}^0\langle\B|\widehat V|\C\rangle^0 \over E_\A-E_\C+i\epsilon} \widehat {\cal T}(\C|\A),
\fe
where the measure $d\C$ is normalized such that $\int d\C |\C\rangle^0\, {}^0\langle \C|=1$.

Now we will restrict to the 2-particle sector, where $\widehat V$ can be replaced with $\widehat{\cal H}_4$. Let $\overline p_\A \equiv (p_\A^+, p_\A^\perp)$ be the lightfront momentum of $|\A\rangle^0$. Due to momentum conservation, matrix elements of $\widehat\cH_4$ take the form
\ie\label{overlapnotation}
{}^0\langle\B| \widehat\cH_4 |\A\rangle^0 \equiv (2\pi)^2 \delta^2(\overline p_\B - \overline p_\A) {\cal H}_4(\B|\A).
\fe
We will also define the reduced $T$-matrix elements ${\cal T}(\B|\A)$ by
\ie
\widehat {\cal T}(\B|\A) \equiv (2\pi)^2 \delta^2(\overline p_\B - \overline p_\A) {\cal T}(\B|\A).
\fe
The equation (\ref{hattintegraleqn}) can be reduced to
\ie\label{highleveleqn}
{\cal T}(\B|\A) = {\cal H}_4(\B|\A) + \int d\C\, (2\pi)^2 \delta^2(\overline p_\C - \overline p_\A) { {\cal H}_4(\B|\C)\over E_\A-E_\C+i\epsilon} {\cal T}(\C|\A).
\fe
Note that in this form of the scattering equation, the lightfront momentum conservation $\overline p_\B=\overline p_\A$ is always imposed, whereas energy conservation is not enforced.

Next, we specialize to the 2-particle gauge singlet sector, spanned by the basis states
\ie
\frac{1}{\sqrt{N}}\,a_{\bar i}^\dagger(\overline p_1) b_i^\dagger(\overline p_2) |0\rangle.
\fe
Without loss of generality, we can restrict to the sector with fixed total lightcone momentum $P^+$, and transverse momentum $P^\perp=0$. We will further adopt the convention (\ref{momentasplit}) and label the basis states by the transverse momentum $p$ and the fraction $x$ of lightcone momentum shared by one of the particles, 
\ie\label{freestates}
|x,p\rangle^0 \equiv \frac{1}{\sqrt{N}}\,a_{\bar i}^\dagger\left((1-x)P^+, -p\right) b_i^\dagger\left(xP^+, p\right)  |0\rangle.
\fe
The $\widehat\cH_2$-eigenvalue of the state (\ref{freestates}), as follows from (\ref{eq:H2}), is
\ie\label{H2xy}
E(x,p) = {p^2+m^2 \over 2P^+ x(1-x)}.
\fe
The matrix elements of $\widehat\cH_4$, defined as in (\ref{overlapnotation}), are given by (\ref{eq:H4}) with the substitution of variables (\ref{momentasplit}), 
\ie\label{H4elementsxy}
\cH_4(y,q|x,p) = \cH_4^{\rm reg}(y,q|x,p;\delta,\Lambda) = \cH_4^{\rm cl}(y,q|x,p) +  {\cal C}_4 (y,q|x,p;\delta,\Lambda),
\fe
where the two terms on the RHS are the same as those appearing in (\ref{eq:H4}) with the substitution (\ref{momentasplit}). Explicitly, we have
\ie\label{explicitCounterTerms}
& \cH_4^{\rm cl}(y,q|x,p)=- {\pi\lambda\over (P^+)^2}\Bigg\{ m\bigg[\frac{1}{x y}+\frac{1}{(1-x) (1-y)}\bigg] + i {p y(1-y)+q x(1-x) \over x(1-x) y (1-y) (x-y)_\eps} \Bigg\} ,
\\
& {\cal C}_4 (y,q|x,p;\delta,\Lambda) = {\lambda^2\over 2(P^+)^2} \int_{\delta\over P^+}^\infty {dz\over z} \int_{-\Lambda}^\Lambda d\ell\,\bigg[\frac{1}{(1+z) (1-x+z) (1-y+z)}\\
 &\quad\quad\quad\quad\quad\quad\quad\quad\quad\quad\quad\quad\quad\quad\quad\quad\quad\quad\quad\quad\quad\quad\quad\quad-\frac{1}{(z-1)_\eps (z-x)_\eps (z-y)_\eps}\bigg].
\fe
 The equation (\ref{highleveleqn}) for the $T$-matrix elements ${\cal T}(y,q|x,p)$ defined with respect to the basis (\ref{freestates}) and its in-state analog can be written as
\ie\label{eq:intEqT1_unregularized}
\mathcal{T}(y,q|x,p) &= \lim_{\Lambda \to \infty} \lim_{\delta \to 0^+} \Bigg[ \cH_4^{\rm reg}(y,q|x,p;\delta,\Lambda)
\\
&~~~~~ +
{P^+\over (2\pi)^2} \int_{\delta\over P^+}^{1-{\delta\over P^+}} dz \int_{-\Lambda}^{\Lambda} d\ell\,
{\cH_4^{\rm reg}(y,q|z,\ell;\delta,\Lambda) 
\over
\frac{p^2+m^2}{2P^+ x (1-x)} - \frac{\ell^2+m^2}{2P^+ z (1-z)} + i\epsilon}\, \mathcal{T}(z,\ell | x,p)  \Bigg],
\fe
where the intermediate state has lightfront momentum assignment $\overline q_1 = ((1-z) P^+,-\ell)$, $\overline q_2 = (zP^+, \ell)$ for the particle and anti-particle respectively.  

\subsection{Removing the regulators}
\label{sec:bestinteqn}

The equation (\ref{eq:intEqT1_unregularized}) is somewhat awkward to work with as the counter terms appearing in the integration kernel are divergent in the limit $\delta\to 0^+$, $\Lambda\to \infty$. To get a handle on the solution, let us expand the $T$-matrix element as a power series in $\lambda$,
\ie\label{taylorexp}
{\cal T}(y,q|x,p) = \sum_{L=0}^\infty \lambda^{L+1} {\cal T}^{(L)}(y,q|x,p).
\fe
Note that ${\cal H}_4^{\rm reg}$ (\ref{H4elementsxy}) contains an order $\lambda$ classical term and an order $\lambda^2$ counter term. The order $\lambda$ part of (\ref{eq:intEqT1_unregularized}) gives the tree-level amplitude
\ie\label{treelevel}
\lambda {\cal T}^{(0)}(y,q|x,p) = {\cal H}_4^{\rm cl}(y,q|x,p).
\fe 
At order $\lambda^2$, (\ref{eq:intEqT1_unregularized}) gives, after substituting ${\cal H}_4^{\rm cl}$ with $\lambda {\cal T}^{(0)}$, the 1-loop amplitude
\ie\label{oneloopt}
\lambda ^2\,\tT^{(1)}(y,q|x,p) &=  \lim_{\Lambda \to \infty}\lim_{\delta \to 0}\Bigg[ {\cal C}_4(y,q|x,p;\delta,\Lambda) 
\\
&~~~~~~~~ + \lambda^2 {P^+\over (2\pi)^2}\,\int_{
{\delta\over P^+}}^{1- {\delta\over P^+}} dz \int_{-\Lambda}^\Lambda d\ell\,\,  {\tT^{(0)}(y,q|z,\ell)\,\tT^{(0)}(z,\ell | x,p) \over {p^2+m^2\over 2 P^+ x (1-x)} - {\ell^2+m^2\over 2 P^+ z (1-z)} +i\epsilon}\Bigg] .
\fe
The divergent terms in the bracket on the RHS take the form\footnote{The principal value prescription of the $z$-integral arises from the $\varepsilon$-regulator in (\ref{H4explicit}) or (\ref{explicitCounterTerms}).}
\ie\label{c4cancel}
& {\cal C}_4(y,q|x,p;\delta,\Lambda) -\lambda^2 {\Lambda (P^+)^2\over \pi^2} \dashint_{\frac{\delta}{P^+}}^{1-\frac{\delta}{P^+}} dz \, z(1-z) \overline\cT^{(0)}_\infty(y|z) \cT^{(0)}_\infty(z|x) \equiv  \delta_C(y,q|x,p;\delta,\Lambda) ,
\fe
where we have defined
\ie{}
& \cT^{(0)}_\infty(z|x)  \equiv \lim_{\ell \to \infty} { \cT^{(0)}(z,\ell|x,p) \over \ell} = -{\pi i\over (P^+)^2} \frac{1}{  (z-1) z (z-x)},
\\
& \overline\cT^{(0)}_\infty(y|z) \equiv \lim_{\ell \to \infty} { \cT^{(0)}(y,q|z,\ell) \over \ell} =  {\pi i\over (P^+)^2} \frac{1}{  (z-1) z (z-y)} .
\fe
One can verify that
\ie
\lim_{\Lambda \to \infty}\lim_{\delta \to 0}\, \delta_C(y,q|x,p;\delta,\Lambda) = 0,
\fe
and thus (\ref{oneloopt}) is well-defined.

In appendix \ref{app:finite}, we extend this analysis of cancelation of divergence to all orders in $\lambda$. The result is an equivalent, but manifestly finite, integral equation
\ie\label{eq:intEqT1}
\mathcal{T}(y,q|x,p) &=  \cH_4^{\rm cl} (y,q |x, p)
+ \dashint_{0}^1 dz \int_{-\infty}^{\infty} d\ell \left[ {(P^+)^2\over 4\pi^2} { \cH_4^{\rm cl}(y,q|z,\ell) \mathcal{T}(z,\ell |x,p) \over \frac{p^2+m^2}{2 x (1-x)} - \frac{\ell^2+m^2}{2 z (1-z)} + i\epsilon} + {\lambda\over 2\pi i} {\mathcal{T}_\infty(z|x,p)\over z-y} \right] ,
\fe
where ${\cal T}_\infty$ is defined as
\ie
\label{eq:Tinfty}
\mathcal{T}_\infty(z|x, p) \equiv \lim_{\ell \to \infty} {\mathcal{T}(z,\ell | x,p)\over \ell}.
\fe
The $\ell$ integral on the RHS of (\ref{eq:intEqT1}) is defined in the sense of principal value: we place a symmetric cutoff $-\Lambda<\ell<\Lambda$ and take the limit $\Lambda\to \infty$. The $z$-integral is also defined by the principal value prescription that regularizes poles in $z$ along the interval $[0,1]$.

Recall that the $(z,\ell)$-integral originates from the summation over intermediate 2-particle states. The total lightfront momentum is fixed to be $(P^+,P^\perp=0)$, but the energy $H=-P^-$ of the intermediate 2-particle state is unconstrained in the Lippmann-Schwinger formalism. On the other hand, as we explain below, the $\ell$-integral can be reduced to residue contributions from poles at which energy conservation is obeyed.

It is evident from (\ref{eq:intEqT1}) that the solution $\mathcal{T}(y,q|x,p)$ must be linear in $q$.\footnote{We caution that the coefficient of $q$ in $\cT(y,q|x,p)$ diverges at $y=x_*^\pm$, which leads to a discontinuity at zero angle in a sense explained in section \ref{sec:disc}.} It follows that the $\ell$-integrand falls off like $\ell^{-1}$ at infinity, and we can evaluate the principal value $\ell$-integral by averaging between two contours, one enclosing the residue at $\ell=\ell_*(z)$ on the upper half complex $\ell$-plane in the counterclockwise direction, the other enclosing the residue at $\ell=-\ell_*(z)$ on the lower half plane in the clockwise direction, with
\ie\label{ellresidue}
\ell_*(z) = \sqrt{s z(1-z) - m^2+i\epsilon}.
\fe
Here the branch of the square root is chosen such that ${\rm Im}(\ell_*(z))>0$. The Mandelstam variable $s$ is related to $x,p$ by $s = 2P^+ E(x,p) = {p^2+m^2\over x(1-x)}$.
Note that in the limit $\epsilon\to 0^+$, $\ell_*(z)$ either takes positive real value or becomes purely imaginary with a positive imaginary part. 

\begin{figure}[h!]
	\def\svgwidth{1\linewidth}
	\centering{
		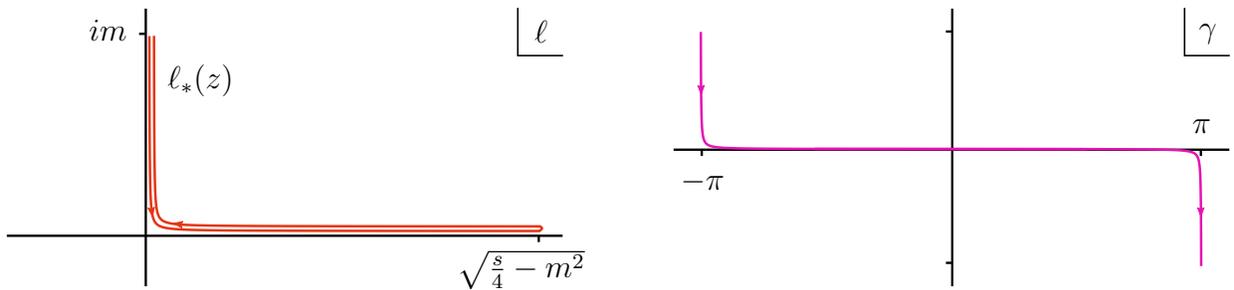	\caption{Left panel: path of $\ell_*(z)$ in the complex $\ell$-plane as $z$ varies from 0 to 1. Right panel: path of the intermediate state's angle $\gamma$ as $z$ ranges from 1 to $z_*^- =  (\sqrt{s}-\sqrt{s-4m^2})/(2\sqrt{s})$ and then back. 
			\label{ellplot}
	}}
\end{figure}

Expressing $p,q$ in terms of $x,y$ and $s$ via (\ref{pqsrel}), we can repackage the $T$-matrix element {\it at energy-conserving kinematics} as
\ie\label{eq:Tab}
{\cal T}^{ba}(y|x) \equiv {\cal T}\big(y, q= b \ell_*(y) \big| x, p = a \ell_*(x) \big),~~~~a,b=\pm.
\fe
The dependence on $s$ is implicit. At a given physical value of $s$ ($>4m^2$), $x$ is assumed to take value in the physical region $[x_*^-, x_*^+]$ as shown in Figure \ref{anglechanging}. We will, however, extend the definition (\ref{eq:Tab}) to $y\in (0,1)$, which includes unphysical values of $y$ outside of the interval $[x_*^-, x_*^+]$ corresponding to complex scattering angles, using the linearity of $\cT(y,q|x,p)$ in $q$ (before imposing energy-conservation).\footnote{We caution the reader that $T^{+a}(y|x)$ and $T^{-a}(y|x)$ do not agree at $y=x_*^\pm$, in particular, due to the singular $q$-dependence of $\cT(y,q|x,p)$ at outgoing angle $0$ or $\pi$.}

After performing the $\ell$-integral on the RHS of (\ref{eq:intEqT1}), the integral equation can be expressed in terms of the energy-conserving amplitudes (\ref{eq:Tab}) in the form (\ref{shortbestintegraleqn}), 
\ie\label{bestintegraleqn}
\boxed{
~\,\mathcal{T}^{ba}(y|x) =  \cV^{ba}(y|x) + {(P^+)^2 \over 4\pi i} \dashint_0^1 dz \sum_{c=\pm} {z(1-z)\over \ell_*(z)}  \cV^{bc}(y|z) \mathcal{T}^{ca}(z|x)~ }
\fe
with 
\ie\label{vbaexpr}
	& {\cal V}^{ba}(y|x) = \cH_4^{\rm cl}\big(y, q= b \ell_*(y) \big| x, p = a \ell_*(x) \big)  .
\fe
The principal value prescription in the $z$-integral arises after taking the $\varepsilon\to 0$ limit on the propagators appearing in ${\cal V}^{ba}$ (see (\ref{explicitCounterTerms})).

We emphasize that the amplitude $\cT^{ba}(y|x)$ appearing (\ref{bestintegraleqn}) is defined beyond the physical kinematic domain, and agrees with the physical amplitude when restricted to $y\in [x_*^-, x_*^+]$. This notion of extended amplitude, as explained below (\ref{eq:Tab}), does not assume analyticity in $y$. In fact, as already mentioned, $\cT^{ba}(y|x)$ is not a function in $y$ but a distribution that contains singular support at forward angle.

\subsection{Restoring Lorentz invariance}
\label{sec:restlor}

Passing from lightcone coordinates to the standard Minkowskian coordinates $(x^0, x^1, x^2)$, the lightfront momenta (\ref{momentasplit}) of the asymptotic particles correspond to the spatial momenta $\vec p_i=(p_i^1, p_i^2)$, with
\ie\label{comassignment}
& \vec p_1 = \left(|\vec p| \cos\A,  |\vec p| \sin\A \right), ~~~~ \vec p_2 = \left( -|\vec p| \cos\A,- |\vec p| \sin\A \right),
\\
& \vec p_3 = \left( |\vec p| \cos\B,  |\vec p| \sin\B \right), ~~~~ \vec p_4 = \left( - |\vec p| \cos\B, - |\vec p| \sin\B \right),
\fe
where $|\vec p| = {1\over 2} \sqrt{s-4m^2}$. The 2-particle in-state, denoted $|x,p\rangle^{\rm in}$ in section \ref{sec:lseqn}, can alternatively be parameterized by the angle $\A$ (at given $s$), related to $x$ and $p$ by (\ref{eq:xtoangle}). The out-state $|y,q\rangle^{\rm out}$ can be parameterized by $\B$ analogously.

In the covariant formulation of scattering amplitudes, the 1-particle state $|\vec p\rangle$ is defined with the normalization $\langle \vec p|\vec p'\rangle =2\sqrt{\vec p^2+m^2} (2\pi)^2 \delta^2(\vec p-\vec p')$, and such that \cite{Weinberg:1995mt}
\ie
|\vec p\rangle = U(L(\vec p))|\vec 0\rangle, 
\fe
where $L(\vec p)$ is a Lorentz boost that takes the particle at rest to one with spatial momentum $\vec p$, and $U(L(\vec p))$ is the corresponding unitary operator. For given $\vec p$, $L(\vec p)$ is specified up to the right-multiplication by an arbitrary spatial rotation, which amounts to a phase ambiguity in the definition of the 1-particle state. The 2-particle asymptotic states appearing in the Lorentz invariant S-matrix element (\ref{sanygen}) are such that $|\vec p_1, \vec p_2\rangle^{\rm in}$ with the assignment (\ref{comassignment}) is related to the 2-particle in-state at zero angle by acting with $U(R(\A))$, where $R(\A)$ is the spatial rotation by $\A$. We expect such basis states to differ from $|x,p\rangle^{\rm in}$ defined in the lightfront quantization by
\ie\label{xpphase}
|x,p\rangle^{\rm in} = {\cal N}(\A) |\vec p_1, \vec p_2\rangle^{\rm in},
\fe
and similarly for the out-states
\ie\label{yqphase}
|y,q\rangle^{\rm out} = {\cal N}(\B) |\vec p_3, \vec p_4\rangle^{\rm out},
\fe 
where the factor ${\cal N}(\A)$ depends nontrivially on the angle $\A$, and takes the form
\ie\label{normi}
{\cal N}(\A) = \left|\cN(\alpha)\right|\, e^{i \varphi(\A)}.
\fe
In particular, its norm $|{\cal N}(\A)|$ is determined by the normalizations of the basis states to be
\ie{}
\left|\cN(\alpha)\right| = \frac{1}{\sqrt{2 s(1-x)x}} =\frac{1}{2\sqrt{p_1^+ p_2^+ }} = \sqrt{2\over s - (s-4m^2) \cos^2\A}.
\fe
The phase $\varphi(\A)$ will be determined in section \ref{minconjecture}. In particular, we will see that it contains an analytic dependence in $\A$ that can be understood through nontrivial Lorentz rotations relating the different asymptotic particle basis states, and a discontinuity at $\A=0,\pi$ due to the choice of lightcone gauge. (\ref{xpphase}) and (\ref{yqphase}) then lead to the relation (\ref{ttmatch}) between ${\cal T}^{ba}(y|x)$ appearing in the equation (\ref{bestintegraleqn}) and the covariant singlet channel amplitude $T_{\rm S}(s,\theta)$ of (\ref{sanygen}).

\section{Solving the scattering equation}
\label{sec:solution}

To find the solution to (\ref{bestintegraleqn}), we shall assume the ansatz (\ref{eq:genform}) which will be justified a posteriori. Namely, the amplitude $\cT^{ba}(y|x)$ is the sum of a function $\widetilde\cT^{ba}(y|x)$ and a distribution supported in the forward direction of the form $\delta(x-y)\delta^{ba} \cT_0(x)$. We will refer to them as the ``function part" and the ``forward-distribution part" of the amplitude, respectively. Both can be expanded as power series in the 't Hooft coupling,
\ie\label{taylorexpCurly}
& \widetilde{\cal T}^{ba}(y|x) = \sum_{L=0}^\infty \lambda^{L+1} \widetilde{\cal T}^{ba,(L)}(y|x),
\\
&  \cT_0(x) =  \sum_{L=0}^\infty \lambda^{L+1} \cT_0^{(L)}(x),
\fe
and analyze order by order before identifying the full solution. 

We will also assume that the phase in $\varphi(\A)$ appearing (\ref{normi}) has the expansion
\ie\label{eqn:phase}
\varphi(\alpha) \equiv \sum_{L=0}^\infty \lambda^{L} \varphi^{(L)}(\alpha) .
\fe
Note that the leading phase correction $\varphi^{(0)}(\A)$ is independent of $\lambda$.
The analogous expansion of the covariant amplitude (\ref{ttapp}) takes the form
\ie{}
& \widetilde T(s,\theta) = \sum_{L=0}^\infty \lambda^{L+1} \widetilde T^{(L)}(s,\theta),
\\
&  T_0(s) =  \sum_{L=0}^\infty \lambda^{L+1} T_0^{(L)}(s),
\fe

\subsection{Tree and one-loop results}
\label{sec:treeoneloop}

The tree-level amplitude given by (\ref{treelevel}) has no forward-distribution part, namely $\cT_0^{(0)}(x)=0$ or equivalently $T_0^{(0)}(s)=0$. Its function part is
\ie\label{treetildt}
\widetilde\cT^{ba,(0)}(y|x) = \lambda^{-1}\cV^{ba}(y|x),
\fe
with $\cV^{ba}$ defined by (\ref{vbaexpr}), (\ref{explicitCounterTerms}).
The corresponding tree-level covariant amplitude, related by (\ref{ttmatch}), (\ref{normi}), is
\ie\label{treesoln}
\widetilde T^{(0)}(s,\theta=\B-\A) &= {e^{i\varphi^{(0)}(\beta)-i\varphi^{(0)}(\A)} \over \big|\cN(\B) \cN(\A)\big|} \widetilde\cT^{ba,(0)}(y|x)
\\
&= 8\pi m - 4\pi i\sqrt{s}\cot{\beta-\alpha\over 2} ,
\fe
where the leading phase correction $\varphi^{(0)}(\A)$ is determined, up to linear terms in $\A$, by the requirement that (\ref{treesoln}) depends only on the difference between the angles $\A$ and $\B$. The result is
\ie\label{eq:0thorderphase}
\varphi^{(0)}(\alpha) = - {i\over 2} \log {2 m + i \sqrt{s} \tan\alpha \over 2 m - i \sqrt{s} \tan \alpha}  =-{i\over 2} \log{ u_a^+(x)\over u_a^-(x)} ,
\fe
where for later convenience we have defined
\ie\label{uadef}
u_a^\pm (x)\equiv { m(1-2x) \mp i a \ell_*(x) \over x(1-x)}.
\fe
In (\ref{eq:0thorderphase}), the branch of the logarithm is chosen such that $\varphi^{(0)}(\A)$ varies continuously from $-\pi$ to $\pi$, as $\A$ ranges from $-\pi$ to $\pi$. In particular, $\varphi^{(0)}(\A)$ vanishes at $\A=0$, corresponding to $x=x_*^-$. The physical origin of the angular dependence of (\ref{eq:0thorderphase}) will be explained in section \ref{minconjecture}.

Next, we consider the one-loop amplitude obtained by inserting the tree-level result ${\cal T}^{ca,(0)}(z|x)$ into the RHS of (\ref{bestintegraleqn}),
\ie\label{tonelooptt}
\cT^{ba,(1)}(y|x) &\equiv \delta(x-y) \delta^{ba} {\cal T}_0^{(1)}(x) + \widetilde\cT^{ba,(1)}(y|x)
\\
&= {(P^+)^2 \over 4\pi i} \dashint_0^1 dz \sum_{c=\pm} {z(1-z)\over \ell_*(z)}  \cT^{bc,(0)}(y|z) \cT^{ca,(0)}(z|x).
\fe
It is convenient to analytically extend $\cT^{ba,(0)}(y|x)$ off the real $x$- and $y$-axis, and view the $z$-integral on the RHS of (\ref{tonelooptt}) as a contour integral.
We begin by analyzing the singularity in the limit $y\to x$, where a pair of poles of $\cT^{bc,(0)}(y|z)$ and $\cT^{ca,(0)}(z|x)$ pinch the $z$-integration contour. In the case $b=-a$, corresponding to backward scattering, one can verify that one of these poles is canceled against a vanishing numerator, leaving a finite result. A singular behavior occurs in the $b=a$ case, corresponding to forward scattering, where the RHS of (\ref{tonelooptt}) is dominated by the contribution from $c=a$, $z\sim x\sim y$,
\ie\label{tildetone}
\cT^{aa,(1)}(y|x) &\sim - {\pi i\over (P^+)^2} { \ell_*(x) \over x(1-x) } \int {dz \over (z-x)_\varepsilon(z-y)_\varepsilon}
\\
& = - {\pi^3 i\over (P^+)^2} { \ell_*(x) \over x(1-x) } \delta(x-y),
\fe
giving the forward-distribution part of (\ref{tonelooptt}) as anticipated. The corresponding term in the covariant amplitude is\footnote{Note that the phase correction in the relation (\ref{ttmatch}) cancels in the forward limit $\B\to \A$.}
\ie
T_0^{(1)} (s) =  -4\pi^3 i \sqrt{s}.
\fe
Away from the forward limit, we integrate the second line of (\ref{tonelooptt}) to find the one-loop amplitude
\ie\label{oineloopres}
\widetilde\cT^{ba,(1)}(y|x) = I_{\rm phys}^{ba}(y|x) + I_{\rm unphys}^{ba}(y|x),
\fe
where
\ie\label{iphysun}
& I_{\rm phys}^{ba}(y|x) = {\pi\over 2 (P^+)^2 } {u_a^+(x) u_b^-(y)\over \sqrt{s}}  \pi i ,
\\
&  I_{\rm unphys}^{ba}(y|x) = {\pi\over 2 (P^+)^2 } \Bigg\{ {u_a^+(x) u_b^-(y)\over \sqrt{s}} \log \frac{\sqrt{s}+2m}{\sqrt{s}-2m} 
\\
&~~~~~~~~~~~~~~~~~~~~~~~~~~~~
+ {u_a^+(x) - u_b^-(y)\over x-y} \left[ {i\pi\over 2} (b-a) + \log{u_a^-(x) u_b^+(y)\over u_a^+(x) u_b^-(y)} \right] \Bigg\}
\fe
are the contributions from the integration over the physical region $z\in (x_*^-,x_*^+)$ (corresponding to positive real $\ell_*(z)$) and the unphysical region $z\in (-1,x_*^-) \cup (x_*^+,1)$ (corresponding to positive imaginary $\ell_*(z)$) respectively. The functions $u_a^\pm(x)$ are defined as in (\ref{uadef}).

The logarithm appearing in the second term in the bracket of (\ref{iphysun}) is defined with the same choice of branch as in (\ref{eq:0thorderphase}) for $x,y$ in the physical region $[x_*^-,x_*^+]$, and analytically continued to $(0,1)$ using the expression (\ref{ellresidue}) for $\ell_*$ with $i\epsilon$ prescription. A consequence of this branch structure is that $I^{+, a}_{\rm unphys}(y|x)$ and $I^{-, a}_{\rm unphys}(y|x)$ do not agree in the limit $y\to x_*^\pm$ or $\ell_*(y)\to 0$, corresponding to outgoing angle $\B=0,\pi$, and likewise $I^{b,+}_{\rm unphys}(y|x)$ and $I^{b,-}_{\rm unphys}(y|x)$ do not agree in the limit $x\to x_*^\pm$ or $\ell_*(x)\to 0$, corresponding to incoming angle $\A=0,\pi$.

The one-loop contribution to the analytic part of the covariant amplitude is thus given by
\ie\label{notilde}
\widetilde T^{(1)}(s,\theta=\B-\A) 
& = { e^{i\varphi^{(0)}(\beta)-i\varphi^{(0)}(\A)} \over \big|\cN(\B) \cN(\A)\big|} \cT^{ba,(1)}(y|x)  +  i\left[ \varphi^{(1)}(\B)-i\varphi^{(1)}(\A) \right] \widetilde T^{(0)}(s,\B-\A) 
\\
&= 2\pi\,{s-4m^2\over \sqrt{s}} \left( \log {\sqrt{s}+2m\over \sqrt{s}-2m} + \pi i \right) .
\fe
In deriving the last equality, the phase correction $\varphi^{(1)}(\A)$ is again fixed, up to linear terms in $\A$, by demanding that $\widetilde T^{(1)}$ depends on $\A,\B$ only through $\theta=\B-\A$, giving the curious-looking result
\ie\label{phicur}
\varphi^{(1)}(\alpha) = \varphi^{(0)}(\alpha) + \pi\, \Theta(-\A),~~~~-\pi<\A<\pi,
\fe
where $\Theta$ is the Heaviside step function. The appearance of the discontinuity in $\A$ is a consequence of the branch structure of (\ref{iphysun}) as described below (\ref{uadef}). The physical origin of (\ref{phicur}) will be explained in section \ref{minconjecture}.

\subsection{Forward singularity to all orders}
\label{sec:forward}

Up to one-loop order, we have encountered two types of singularities in the forward scattering limit $\theta\to 0$, in both the tree amplitude $\widetilde T^{(0)}(s,\theta)$ which has a pole at $\theta=0$, and the 1-loop amplitude $T^{(1)}(s,\theta)$ which contains the forward-distribution $\delta(\theta) T_0^{(1)}(s)$ supported at $\theta=0$. We will now argue that these are the only types of singularities that can arise in the forward limit at any order in $\lambda$, namely
\ie\label{ansatz}
\mathcal{T}^{ba}(y|x) = f(\lambda) \widetilde\cT^{ba,(0)}(y|x) + g(\lambda)\,\delta(x-y)\delta^{ab} \cT_0^{(1)}(x) +({\rm non-singular}),
\fe 
for some functions $f(\lambda)=\lambda + {\cal O}(\lambda^3)$, $g(\lambda)=\lambda^2 + {\cal O}(\lambda^4)$.

Indeed, the tree amplitude $\widetilde\cT^{ba,(0)}(y|x)$ given by (\ref{treetildt}), and the 1-loop forward-distribution coefficient ${\cal T}_0^{(1)}(x) = - {\pi^3 i\over (P^+)^2} {\ell_*(x)\over x(1-x)}$ as in (\ref{tildetone}), obey the relation
\ie\label{feeda}
& {(P^+)^2\over 4\pi i} \dashint_0^1 dz \sum_{c=\pm} {z(1-z)\over \ell_*(z)}  \cV^{bc}(y|z) \widetilde\cT^{ca,(0)}(z|x) = \lambda\, \delta(x-y)\delta^{ab} \cT_0^{(1)}(x) + ({\rm non-singular}),
\fe
as already seen in section \ref{sec:treeoneloop}, as well as
\ie\label{feedb}
& {(P^+)^2\over 4\pi i} \dashint_0^1 dz \sum_{c=\pm} {z(1-z)\over \ell_*(z)}  \cV^{bc}(y|z) \delta(z-x) \delta^{ca} \cT_0^{(1)}(x) = - {\pi^2\lambda\over 4} \widetilde\cT^{ba,(0)}(y|x) ,
\fe
which gives a 2-loop singularity of the same type as one that occurs at tree-level. This cyclic pattern persists to all orders. Plugging the ansatz (\ref{ansatz}) into (\ref{bestintegraleqn}), and using (\ref{feeda}), (\ref{feedb}), we deduce that $f(\lambda)$ and $g(\lambda)$ obey the recursive relations
\ie\label{geqn}
f(\lambda) = \lambda -\frac{\pi^2\lambda}{4 }g(\lambda),~~~~~ g(\lambda) = \lambda f(\lambda),
\fe
from which we solve
\ie\label{feqn}
f(\lambda) = {\lambda\over 1+{\pi^2\over 4} \lambda^2}.
\fe 
This in particular determines the forward-distribution part of the amplitude in the lightfront basis to all orders,
\ie\label{tzerox}
{\cal T}_0(x) = g(\lambda) {\cal T}_0^{(1)}(x).
\fe
(\ref{tzerox}) agrees with $T_0(s)$ of (\ref{ttconjres}) provided the identification
\ie
g(\lambda) = {2\over \pi^2} \left(1-\cos(\pi\widetilde\lambda) \right).
\fe
This is equivalent to the coupling redefinition (\ref{couplingredef}), and $f(\lambda)={1\over\pi}\sin(\pi\widetilde\lambda)$.

\subsection{A monodromy and a discontinuity}
\label{sec:disc}

\begin{figure}[h!]
	\def\svgwidth{.7\linewidth}
	\centering{
\begingroup%
  \makeatletter%
  \providecommand\color[2][]{%
    \errmessage{(Inkscape) Color is used for the text in Inkscape, but the package 'color.sty' is not loaded}%
    \renewcommand\color[2][]{}%
  }%
  \providecommand\transparent[1]{%
    \errmessage{(Inkscape) Transparency is used (non-zero) for the text in Inkscape, but the package 'transparent.sty' is not loaded}%
    \renewcommand\transparent[1]{}%
  }%
  \providecommand\rotatebox[2]{#2}%
  \newcommand*\fsize{\dimexpr\f@size pt\relax}%
  \newcommand*\lineheight[1]{\fontsize{\fsize}{#1\fsize}\selectfont}%
  \ifx\svgwidth\undefined%
    \setlength{\unitlength}{500bp}%
    \ifx\svgscale\undefined%
      \relax%
    \else%
      \setlength{\unitlength}{\unitlength * \real{\svgscale}}%
    \fi%
  \else%
    \setlength{\unitlength}{\svgwidth}%
  \fi%
  \global\let\svgwidth\undefined%
  \global\let\svgscale\undefined%
  \makeatother%
  \begin{picture}(1,0.26)%
    \lineheight{1}%
    \setlength\tabcolsep{0pt}%
    \put(0,0){\includegraphics[width=\unitlength,page=1]{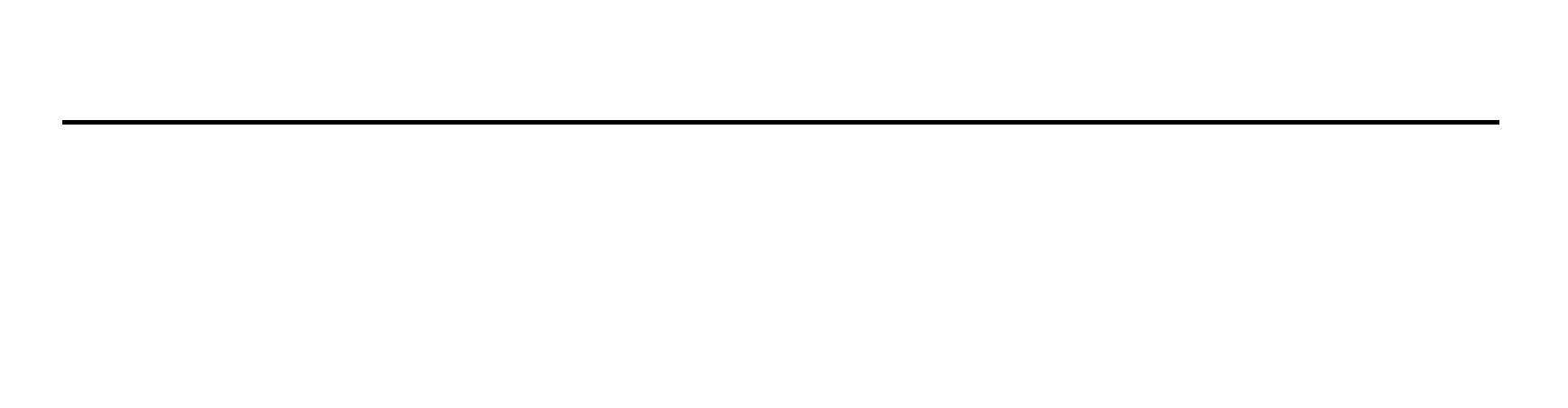}}%
    \put(0.92812756,0.23923396){\makebox(0,0)[lt]{\lineheight{1.25}\smash{\begin{tabular}[t]{l}$z$\end{tabular}}}}%
    \put(0.55597894,0.20780159){\makebox(0,0)[lt]{\lineheight{1.25}\smash{\begin{tabular}[t]{l}$x$\end{tabular}}}}%
    \put(0.555979,0.08011047){\makebox(0,0)[lt]{\lineheight{1.25}\smash{\begin{tabular}[t]{l}$x$\end{tabular}}}}%
    \put(0.4336752,0.20780212){\makebox(0,0)[lt]{\lineheight{1.25}\smash{\begin{tabular}[t]{l}$y$\end{tabular}}}}%
    \put(0,0){\includegraphics[width=\unitlength,page=2]{PV2contour.pdf}}%
    \put(0.92812701,0.10848943){\makebox(0,0)[lt]{\lineheight{1.25}\smash{\begin{tabular}[t]{l}$z$\end{tabular}}}}%
    \put(0,0){\includegraphics[width=\unitlength,page=3]{PV2contour.pdf}}%
    \put(0.22203374,0.08011012){\makebox(0,0)[lt]{\lineheight{1.25}\smash{\begin{tabular}[t]{l}$y$\end{tabular}}}}%
  \end{picture}%
\endgroup%
	\caption{The $\ddashint$ notation represents the average between the two contours, shown in orange and blue respectively.
			\label{PV2contour}
	}}
\end{figure}

It follows from the structure of (\ref{bestintegraleqn}) that the amplitude $\cT^{ba}(y|x)$ is analytic in $y$ away from $y=x$ and away from the branch points $y=x_*^\pm$. We will extend the definition of the function part of the amplitude $\widetilde\cT^{ba}(y|x)$ by analytic continuation from the physical region $y\in (x_*^-,x_*^+)$, and rewrite (\ref{bestintegraleqn}) as an integral equation for $\widetilde\cT^{ba}(y|x)$ in the form
\ie\label{secondbestintegraleqn}
\widetilde\cT^{ba}(y|x) =  {f(\lambda)\over \lambda}\cV^{ba}(y|x) + {(P^+)^2\over 4\pi i} \ddashint_0^1dz \sum_{c=\pm} {z(1-z) \over \ell_*(z)} \cV^{bc}(y|z) \widetilde\cT^{ca}(z|x),
\fe
where the notation $\ddashint$ stands for the average between the integrals over a pair of contours that run above and below the points $z=x$ and $z=y$, along the interval $(0,1)$ in the complex $z$-plane, shown in Figure \ref{PV2contour}.

Recall that the transverse momentum $q$ is related to $y$ and $b=\pm$ by $q(y)= b\ell_*(y)$ in (\ref{eq:Tab}). As $y$ circles around the branch point $y=x_*^-$, $\ell_*(y)$ turns into $\ell_*^\circlearrowleft(y) = -\ell_*(y)$. We denote by $\widetilde\cT^{ba\circlearrowleft}(y|x)$ the analytic continuation of $\widetilde\cT^{ba}(y|x)$ under this monodromy. Naively, analytic continuation of the momenta suggests that $\widetilde\cT^{ba\circlearrowleft}(y|x)$ should be related to $\widetilde\cT^{-b,a}(y|x)$. Their precise relation can be seen as follows.

\begin{figure}[h!]
	\def\svgwidth{.7\linewidth}
	\centering{
\begingroup%
  \makeatletter%
  \providecommand\color[2][]{%
    \errmessage{(Inkscape) Color is used for the text in Inkscape, but the package 'color.sty' is not loaded}%
    \renewcommand\color[2][]{}%
  }%
  \providecommand\transparent[1]{%
    \errmessage{(Inkscape) Transparency is used (non-zero) for the text in Inkscape, but the package 'transparent.sty' is not loaded}%
    \renewcommand\transparent[1]{}%
  }%
  \providecommand\rotatebox[2]{#2}%
  \newcommand*\fsize{\dimexpr\f@size pt\relax}%
  \newcommand*\lineheight[1]{\fontsize{\fsize}{#1\fsize}\selectfont}%
  \ifx\svgwidth\undefined%
    \setlength{\unitlength}{500bp}%
    \ifx\svgscale\undefined%
      \relax%
    \else%
      \setlength{\unitlength}{\unitlength * \real{\svgscale}}%
    \fi%
  \else%
    \setlength{\unitlength}{\svgwidth}%
  \fi%
  \global\let\svgwidth\undefined%
  \global\let\svgscale\undefined%
  \makeatother%
  \begin{picture}(1,0.14)%
    \lineheight{1}%
    \setlength\tabcolsep{0pt}%
    \put(0,0){\includegraphics[width=\unitlength,page=1]{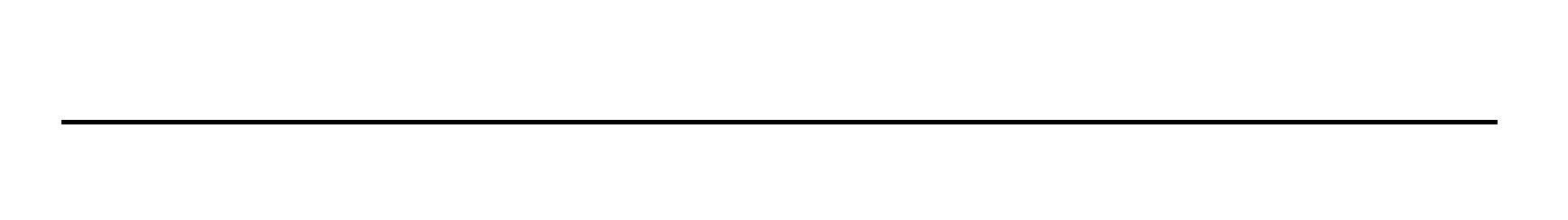}}%
    \put(0.93042078,0.11702813){\makebox(0,0)[lt]{\lineheight{1.25}\smash{\begin{tabular}[t]{l}$z$\end{tabular}}}}%
    \put(0.55625897,0.08480212){\makebox(0,0)[lt]{\lineheight{1.25}\smash{\begin{tabular}[t]{l}$x$\end{tabular}}}}%
    \put(0.43296835,0.08480212){\makebox(0,0)[lt]{\lineheight{1.25}\smash{\begin{tabular}[t]{l}$y$\end{tabular}}}}%
    \put(0,0){\includegraphics[width=\unitlength,page=2]{MonodromyContour.pdf}}%
  \end{picture}%
\endgroup%
	\caption{Analytic continuation of the $T$-matrix element as $y$ moves around the branch point. 
			\label{MonodromyContour}
	}}
\end{figure}

As we analytically continue both sides of the equation (\ref{secondbestintegraleqn}) in $y$, starting from the physical region and around the branch point $y=x_*^-$, $\cV^{bc}(y|z)$ in the integration kernel has a pole at $z=y$ that moves in the complex $z$-plane, crossing the $z$-integration contour twice (Figure \ref{MonodromyContour}). At the first crossing, the $z$-integral picks up a residue contribution from the $c=b$ term, proportional to $\widetilde\cT^{ba\circlearrowleft}(y|x)$.
As we further continue $y$ onto the second sheet, the $z$-integral picks up a second residue contribution from the $c=-b$ term, proportional to $\widetilde\cT^{-b, a}(y|x)$. Together they give rise to the monodromy
\ie\label{apptest3}
\cT^{ba\circlearrowleft}(y|x) &= {f(\lambda)\over \lambda}\cV^{ba\circlearrowleft}(y|x) + {(P^+)^2\over 4\pi i} \ddashint_0^1dz \sum_{c=\pm} {z(1-z) \over \ell_*(z)} \cV^{bc\circlearrowleft}(y|z) \widetilde\cT^{ca}(z|x)
\\
&~~~ - {b\over 2}i \pi \lambda\,\cT^{ba\circlearrowleft}(y|x)- {b\over 2}i \pi \lambda\,\widetilde\cT^{-b,a}(y|x)
\\
&= \left(1-{b\over 2}i \pi \lambda\right)\widetilde\cT^{-b,a}(y|x)- {b\over 2}i \pi \lambda\,\cT^{ba\circlearrowleft}(y|x) ,
\fe 
where we have used $\cV^{bc\circlearrowleft}(y|x)=\cV^{-b,c}(y|x)$. Solving (\ref{apptest3}) yields
\ie\label{jumpform}
\cT^{ba\circlearrowleft}(y|x) = \frac{1- {b\over 2} i\pi \lambda}{1+ {b\over 2}i\pi \lambda}\widetilde\cT^{-b,a}(y|x) = e^{-i b \pi \tilde{\lambda}}\,\widetilde\cT^{-b,a}(y|x) .
\fe
In the second equality, we used the coupling redefinition (\ref{couplingredef}). Remarkably, the monodromy relation (\ref{jumpform}) takes the form of an anyonic phase (see section \ref{minconjecture} for its interpretation).

Further taking the limit $y\to x_*^-$, and using the continuity of $\cT^{ba}(y|x)$ at the branch point, we see that $\cT^{+,a}(y|x)$ and $\cT^{-,a}(y|x)$ differ by a phase $e^{-i\pi\widetilde\lambda}$ at $y=x_*^-$.  A similar relation between $\cT^{b,+}(y|x)$ and $\cT^{b,-}(y|x)$ at $x=x_*^\pm$ can be derived using the fact that $\cT^{ba}(y|x)$ is related to $\cT^{ab}(x|y)$ by complex conjugation together with flipping the sign of $i\epsilon$ in the Lippmann-Schwinger equation. Together they will give rise to a discontinuity in the phase of ${\cal N}(\A)$ appearing in (\ref{ttmatch}), extending the one-loop observation (\ref{phicur}) to all orders in $\lambda$.\footnote{This discontinuity can alternatively be derived by directly evaluating
\ie\label{altderv}
\cT^{+a}(x_*^-|x) - \cT^{-a}(x_*^-|x) &= \lim_{y\to x_*^-} \Big[ \cT(y,\ell_*(y)|x,a \ell_*(x)) - \cT(y, -\ell_*(y)|x,a \ell_*(x)) \Big]
\\
&= 2 \lim_{y\to x_*^-} \ell_*(y) \cT_\infty(y|x, a\ell_*(x)),
\fe
where we recall that $\cT_\infty$ is defined in (\ref{eq:Tinfty}) as the linear coefficient of $\cT(y,q|x,p)$ in $q$. While $\ell_*(x_*^-)=0$, the limit in the second line of (\ref{altderv}) does not vanish because $\cT_\infty(y|x,p)$ is singular at $y=x_*^-$. Taking the $y\to x_*^-$ limit on the $q$-coefficient of (\ref{eq:intEqT1}), the RHS of (\ref{eq:intEqT1}) is dominated by integration near $z=x_*^-$, giving
\ie
	\lim_{y \to x_*^-} \ell_*(y) \cT_\infty(y|x,a\ell_*(x)) = {\pi i \lambda\over 4} \sum_{c=\pm} \cT^{ca}(x_*^-|x),
\fe
which combines with (\ref{altderv}) to give $\cT^{+a}(x_*^-|x)=e^{-i\pi\widetilde\lambda}\cT^{-a}(x_*^-|x)$.}

\subsection{The all-order solution}
\label{minconjecture}

The expectation (\ref{ttmatch}) suggests the ansatz
\ie\label{ansatztildephys}
\widetilde\cT^{ba}(y|x) = e^{i\varphi(\A)-i\varphi(\B)} \big|\cN(\B) \cN(\A)\big| \widetilde T(s,\theta=\B-\A),
\fe
where ${\cal N}(\A)$ is defined as in (\ref{normi}), and $\varphi(\A)$ is of the form (\ref{eqn:phase}). While the LO phase $\varphi^{(0)}(\A)$ and the NLO phase $\varphi^{(1)}$ are given in (\ref{eq:0thorderphase}) and (\ref{phicur}) respectively,  $\varphi^{(L)}(\A)$ for $L\geq 2$ are yet to be determined.

The discussion of section \ref{sec:restlor} suggests that the phase $\varphi(\A)$ is due to the Lorentz rotation relating the 1-particle states in the lightfront basis to the those of the covariant basis, modulo the discontinuity at $\A=0,\pi$ determined by (\ref{jumpform}). The appearance of the phase discontinuity is unsurprising from the perspective of LSZ relation, given that the lightcone gauge condition is equivalent to attaching a semi-infinite Wilson line to each particle that extends in the lightcone direction, and that the causal domain of dependence of the two Wilson lines overlap when the spatial separation of the two particles is at angle $0$ or $\pi$.

Explicitly, we expect
\ie\label{varphipha}
\varphi(\A) = 2 h(\lambda) \varphi^{(0)}(\A) + \pi\widetilde\lambda\,\Theta(-\A),~~~~-\pi < \A < \pi,
\fe
where $h(\lambda)={1+\lambda \over 2} +{\cal O}(\lambda^2)$ (see (\ref{phicur})) is the anyonic spin of the particle. $\varphi^{(0)}(\A)$, as given by (\ref{eq:0thorderphase}), has the following interpretation. Let $L_i(v)$ be the Lorentz boost in the $x^i$ direction $(i=1,2)$ with velocity $v$. For $v=\sqrt{s-4m^2\over s}$ of the particle in the center of mass frame, we have 
\ie\label{llrrel}
L_1(v \cos\A) L_2({v \sin\A\over \sqrt{1-v^2 \cos^2\A}})R(\varphi^{(0)}(\A)) = R(\A) L_1(v) ,
\fe
where $R(\A)$ stands for the spatial rotation by the angle $\A$. While $U(R(\A) L_1(v))$ takes the 1-particle state at rest $|\vec 0\rangle$ to $|\vec p_2\rangle$ in the covariant basis, $U(L_1(v \cos\A) L_2({v \sin\A\over \sqrt{1-v^2 \cos^2\A}}))$ takes $|0\rangle$ to the 1-particle state of momentum $p_2$ in the lightfront basis. The two basis states are related by the rotation by $R(\varphi^{(0)}(\A))$ in the rest frame of the particle, hence the phase difference (\ref{varphipha}).

\begin{figure}[h!]
	\def\svgwidth{0.45\linewidth}
	\centering{
\begingroup%
  \makeatletter%
  \providecommand\color[2][]{%
    \errmessage{(Inkscape) Color is used for the text in Inkscape, but the package 'color.sty' is not loaded}%
    \renewcommand\color[2][]{}%
  }%
  \providecommand\transparent[1]{%
    \errmessage{(Inkscape) Transparency is used (non-zero) for the text in Inkscape, but the package 'transparent.sty' is not loaded}%
    \renewcommand\transparent[1]{}%
  }%
  \providecommand\rotatebox[2]{#2}%
  \newcommand*\fsize{\dimexpr\f@size pt\relax}%
  \newcommand*\lineheight[1]{\fontsize{\fsize}{#1\fsize}\selectfont}%
  \ifx\svgwidth\undefined%
    \setlength{\unitlength}{360bp}%
    \ifx\svgscale\undefined%
      \relax%
    \else%
      \setlength{\unitlength}{\unitlength * \real{\svgscale}}%
    \fi%
  \else%
    \setlength{\unitlength}{\svgwidth}%
  \fi%
  \global\let\svgwidth\undefined%
  \global\let\svgscale\undefined%
  \makeatother%
  \begin{picture}(1,0.50277778)%
    \lineheight{1}%
    \setlength\tabcolsep{0pt}%
    \put(0,0){\includegraphics[width=\unitlength,page=1]{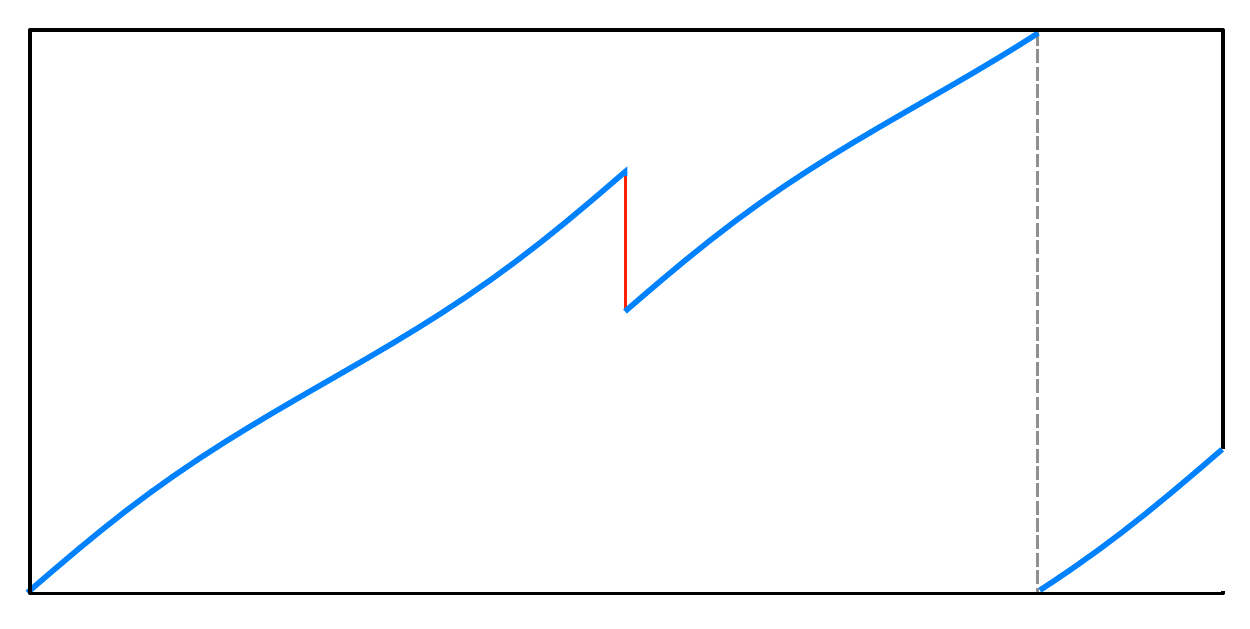}}%
    \put(-0.00748653,-0.01197727){\makebox(0,0)[lt]{\lineheight{1.25}\smash{\begin{tabular}[t]{l}\footnotesize $-\pi$\end{tabular}}}}%
    \put(-0.04889313,0.02955947){\makebox(0,0)[lt]{\lineheight{1.25}\smash{\begin{tabular}[t]{l}\footnotesize $-\pi$\end{tabular}}}}%
    \put(0.96907302,-0.01197595){\makebox(0,0)[lt]{\lineheight{1.25}\smash{\begin{tabular}[t]{l}\footnotesize $\pi$\end{tabular}}}}%
    \put(-0.02389432,0.47068593){\makebox(0,0)[lt]{\lineheight{1.25}\smash{\begin{tabular}[t]{l}\footnotesize $\pi$\end{tabular}}}}%
    \put(0,0){\includegraphics[width=\unitlength,page=2]{phase12.pdf}}%
    \put(0.16603049,0.09866676){\makebox(0,0)[lt]{\lineheight{1.25}\smash{\begin{tabular}[t]{l}\footnotesize $\varphi(\A)$\end{tabular}}}}%
    \put(0.48671368,-0.01197595){\makebox(0,0)[lt]{\lineheight{1.25}\smash{\begin{tabular}[t]{l}\footnotesize $\A$\end{tabular}}}}%
  \end{picture}%
\endgroup%
	\caption{The phase $\varphi(\A)$ mod $2\pi$, given by (\ref{varphipha}) together with (\ref{eq:0thorderphase}), (\ref{anyonspin}), plotted as a function of $\A$ in the case $\widetilde{\lambda}=1/2$. Note the discontinuities at $\A=0$ and $\A=\pi$ marked by the red vertical segments.
			\label{phase12}
	}}
\end{figure}

We propose the following exact formula for the anyonic spin,
\ie\label{anyonspin}
h(\lambda) = {1 + \widetilde\lambda\over 2},
\fe
where $\widetilde\lambda$ is related to $\lambda$ by (\ref{couplingredef}). See Figure \ref{phase12} for the shape of the phase $\varphi(\A)$ as a function of $\A$. Our full analytic ansatz for $\widetilde\cT^{ba}(y|x)$, based on (\ref{ansatztildephys}) with $\widetilde T(s,\theta)$ given by the second line of (\ref{ttconjres}), together with the phase correction (\ref{varphipha}), and the anyonic spin (\ref{anyonspin}), is 
\ie\label{ansatztilde}
& \widetilde\cT^{ba}(y|x) = {2 \sin(\pi\widetilde\lambda) e^{ {i\pi\over 2} \widetilde\lambda (a-b)} \over \sqrt{s x(1-x) y(1-y)}} \left[ {u_a^+(x) u_b^-(y)\over u_a^-(x) u_b^+(y)} \right]^{1+\widetilde\lambda}
\\
& \times \left\{ {1\over \sqrt{s}} \left[ -2m + { x(1-x) y(1-y) \over (x-y) (s-4 m^2)} u_a^-(x) u_b^+(y) (u_a^+(x) - u_b^-(y))\right] - \frac{ 1 + e^{i \pi \widetilde{\lambda } } \left({\sqrt{s}+2 m\over\sqrt{s}-2 m}\right){}^{\widetilde{\lambda }+1}}{1-e^{i \pi \widetilde{\lambda }} \left(\frac{\sqrt{s}+2 m}{\sqrt{s}-2 m}\right){}^{\widetilde{\lambda }+1}}\right\},
\fe
where $u_a^\pm$ is defined in (\ref{uadef}), and the branch is chosen as in (\ref{eq:0thorderphase}). Importantly, (\ref{ansatztilde}) is understood to be defined beyond the physical region $y\in (x_*^-,x_*^+)$, to the entire interval $0<y<1$. This is specified by extending the RHS of (\ref{ansatztilde}) analytically using the expression (\ref{ellresidue}) for $\ell_*$ with $i\epsilon$ prescription, or equivalently, by analytic continuation in $y$ above the branch point $x_*^-$ to $y\in (-1,x_*^-)$, and below the branch point $x_*^+$ to $y\in (x_*^+,1)$.

Indeed, we find that (\ref{ansatztilde}) solves (\ref{secondbestintegraleqn}). This is verified analytically at 2-loop (i.e. $\lambda^3$) order,\footnote{ At 2-loop order, the $y$-dependence of $\widetilde\cT^{ab,(1)}(y|x)$ (\ref{iphysun}) which appears in the RHS of (\ref{secondbestintegraleqn}) makes analytic verification tricky. A simple workaround is to consider another version of the L-S equation with the roles of in- and out- states exchanged, resulting in an equation similar to (\ref{bestintegraleqn}) with $\cT$ and $\cV$ exchanged in the integrand. The average between the two versions of the integration equation is then easy to verify analytically at 2-loop order.}  and numerically at finite $\lambda$, thereby confirming the conjectured result for the singlet sector planar S-matrix element of \cite{Jain:2014nza}. A demonstration of the numerical check is shown in Figure \ref{yplot2}, with further details given in appendix \ref{allorderstest}.

\begin{figure}[h!]
	\def\svgwidth{1\linewidth}
	\centering{
		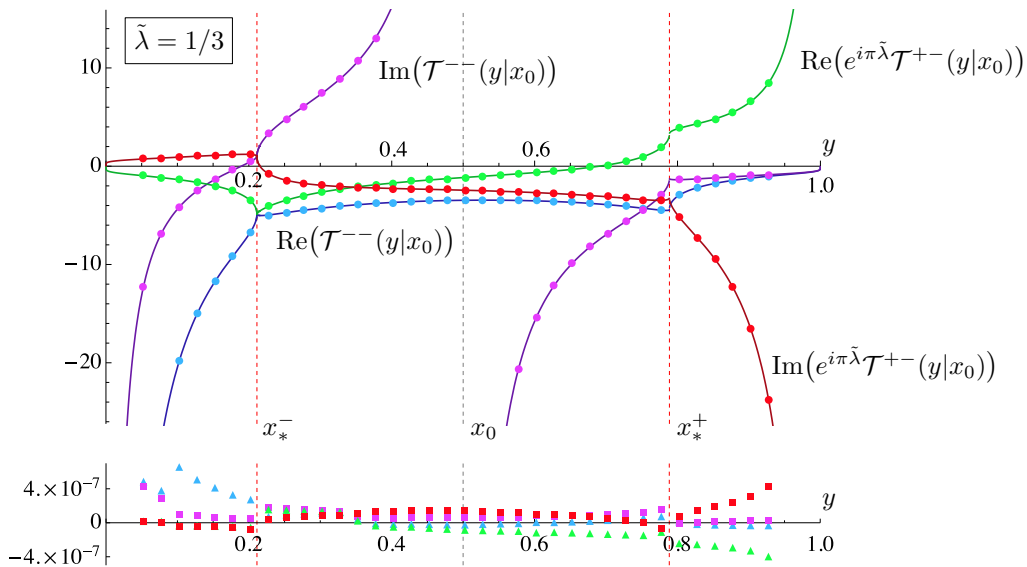	\caption{Numerical verification of the integral equation (\ref{secondbestintegraleqn}) for the ansatz (\ref{ansatztilde}) for $\cT^{ba}(y|x_0)$ with coupling $\widetilde{\lambda}=1/3$, at center-of-mass energy squared $s= 6$ (the physical mass $m$ is set to 1) and incoming angle $\A={\pi\over 2}$ corresponding to $x_0 = {1\over 2}$. The curves in color are given by the analytic ansatz, while the dots are obtained by numerically integrating the RHS of (\ref{secondbestintegraleqn}). The deviation between the two sides of the integral equation, for various components of the amplitude, is shown in the lower panels. As expected, $\cT^{--}(y|x_0)$ agrees with $e^{i\pi\widetilde\lambda} \cT^{+-}(y|x_0)$ at $y=x_*^-$. (Note that $\cT^{--}(y|x_0)$ also agrees with $e^{-i\pi\widetilde\lambda} \cT^{+-}(y|x_0)$ at $y=x_*^+$, not shown in the plot.) }		
	}
\label{yplot2}
\end{figure}

\subsection{Further interpretation of the exact result}

The formula (\ref{anyonspin}) for the spin of the particle, which enters the amplitude in the lightfront basis through the phase correction (\ref{varphipha}), has a nontrivial all-order expansion in $\lambda$. However, it has a simple explanation in terms of the angular dependence of the amplitude, as follows.

Restricted to the 2-particle singlet sector, the planar amplitude (\ref{ttapp}), (\ref{ttconjres}) is related to the partial wave scattering phases $S_\ell(s)$ ($\ell\in\mathbb{Z}$) by
\ie
\delta(\theta) - {i\over 8\pi\sqrt{s}} T_{\rm S}(s,\theta) =  {1\over 2\pi} \sum_{\ell=-\infty}^\infty e^{i\ell\theta} S_\ell(s).
\fe
It follows that the partial wave amplitudes with nonzero angular momentum $\ell$ take a very simple form,
\ie\label{partialwave}
S_\ell(s) &= \int_0^{2\pi} d\theta\, e^{-i\ell\theta} \left[ \cos(\pi\widetilde\lambda) \delta(\theta) - {\sin(\pi\widetilde\lambda)\over 2\pi} \cot{\theta\over 2}\right]
\\
&= e^{ i \pi \widetilde\lambda\,{\rm sign}(\ell)  }.
\fe
At large impact parameter, which corresponds to large $|\ell|$, one expects the nontrivial scattering phase to be entirely due to the anyon statistics, consistent with the correction to spin by precisely ${\widetilde\lambda\over 2}$ as in (\ref{anyonspin}).\footnote{The connection between the spin and statistics of anyons was established in \cite{FROHLICH1991533}.}

Let us also observe that the zeroth partial wave amplitude can be written as
\ie\label{szero}
S_0(s) = {e^{\pi i \widetilde\lambda} z^{\widetilde\lambda+1} - 1\over z^{\widetilde\lambda+1} - e^{\pi i \widetilde\lambda}},~~~~~{\rm where}~ z\equiv {\sqrt{s}-2m\over \sqrt{s}+2m}.
\fe
The simple analytic structure on the complex $s$-plane suggests that (\ref{szero}) may be derived a priori from suitable bootstrap axioms for the anyonic S-matrix.

\section{Discussion}
\label{sec:discuss}

Let us summarize our results so far. Working with an appropriately regularized lightcone Hamiltonian in the Lippmann-Schwinger formulation of scattering theory, we found the exact planar $2\to 2$ S-matrix element in the gauge singlet sector to be in precise agreement with the conjectured result of \cite{Jain:2014nza}, provided that we take into account the coupling redefinition (\ref{couplingredef}), and the phase difference between the bases of asymptotic states according to (\ref{xpphase}), with (\ref{normi}), (\ref{varphipha}), and (\ref{eq:0thorderphase}). The phase factor is tied to the anyonic spin through the Lorentz rotation appearing in (\ref{llrrel}), and is consistent with the anyonic statistics indicated by the partial wave amplitudes (\ref{partialwave}).

The relation (\ref{couplingredef}) between the 't Hooft coupling $\lambda$ in the lightcone Hamiltonian and $\widetilde\lambda$ in the covariant scheme of \cite{Jain:2014nza} should not come as surprise. While $\widetilde\lambda$ ranges over the interval $(-1,1)$ due to the 1-loop renormalization of the Chern-Simons level in the Yang-Mills regularization scheme \cite{Giombi:2011kc, Jain:2014nza}, $\lambda$ can take any real value, consistent with (\ref{couplingredef}).

Let us contrast our analysis with the logic of \cite{Jain:2014nza}. The computation of planar amplitudes in \cite{Jain:2014nza} was based on Feynman diagrams, which in a specific frame can be resummed through Dyson-Schwinger equations in the sector of adjoint, symmetric and antisymmetric representations of the gauge group. The frame choice that made the all-order solution tractable was unavailable for the singlet sector scattering amplitude. Instead, \cite{Jain:2014nza} attempted to obtain the singlet channel amplitude via analytic continuation from other channels formally related by crossing. A naive crossing relation leads to a result that violates unitarity, however. Instead, a modified crossing relation was conjectured, which involves multiplying by a coupling-dependent factor and adding a non-analytic term (the distribution part proportional to $T_0(s)$ in our notation), that led to the conjecture, which we have verified in this paper.

In the lightcone Hamiltonian formalism, the lack of Lorentz symmetry introduced some technical complications, both in terms of the regularization of the Hamiltonian and in the choice of basis of asymptotic states, but they have been overcome in this paper. The payoff is that we have an unambiguous formulation of asymptotic states and the scattering theory. While the computation in this paper is limited to the planar limit, there is no conceptual obstacle in extending this work to subleading orders in $1/N$, and even non-perturbatively at finite $N$ and $k$ (e.g. through Hamiltonian truncation \cite{Delacretaz:2018xbn, Anand:2020gnn, Fitzpatrick:2022dwq}).

The next step we hope to undertake is to obtain the analogous result in the adjoint sector, which should clarify the crossing relation of the $2\to 2$ S-matrix elements of the CSM theory, at least in the planar limit, perhaps eventually at finite $N$. It is also of interest to understand the analyticity property of the S-matrix elements in the lightcone formulation away from the physical domain on general grounds.

\section*{Acknowledgements}

We would like to thank Shiraz Minwalla, Sebastian Mizera and Amit Sever for discussions. JS acknowledges the support of Harvard University, where a portion of this work was completed. BG thanks Tel-Aviv University and the S-matrix Bootstrap IV workshop in Crete for hospitality during the course of this work. This work is supported in part by a Simons Investigator Award from the Simons Foundation, by the Simons Collaboration Grant on the Non-Perturbative Bootstrap, and by DOE grants DE-SC0007870.

\appendix

\section{The lightcone Hamiltonian in terms of ladder operators\label{app:Ham}}
In the main text we decomposed $\widehat{\cal H}_4$ into the pieces giving the tree-level amplitudes in each channel plus the particle number nonconserving interactions,
\ie
\widehat{\cal H}_4 = \widehat{\cal H}^\text{S}_4+\widehat{\cal H}_4^\text{A}+\widehat{\cal H}_4^\text{P-P} +\widehat{\cal H}_4^\text{A-A} +\widehat{\cal H}_4^{3\to 1}+\widehat{\cal H}_4^{1\to 3}\,.
\fe
$\widehat{\cal H}_4^\text{S}$ was given by (\ref{interaction}). The other channels can be worked out similarily,
\ie
\widehat{\cal H}_4^\text{A}&=  \int \prod_{i=1}^4 \frac{d^2\overline p_i}{(2\pi)^2} \Theta(p_i^+)(2\pi)^2 \delta^2(\overline p_{14}+\overline p_{23})\,\cH_4^{\rm reg}(\overline p_1,-\overline p_3,-\overline p_2,\overline p_4; \delta,\Lambda)\,\\&\quad\quad\quad\quad\quad\quad\quad\quad\quad\quad\quad\quad\times b^\dagger_k(\overline p_4) a^\dagger_{\bar \ell}(\overline p_3) b_{\bar k}(\overline p_2) a_\ell(\overline p_1)  \,,\\
\widehat{\cal H}_4^\text{P-P} &=\int \prod_{i=1}^4 \frac{d^2\overline p_i}{(2\pi)^2} \Theta(p_i^+) (2\pi)^2 \delta^2(\overline p_{14}+\overline p_{23})\frac{1}{2}\cH_4^{\rm reg}(\overline p_2,-\overline p_3,\overline p_4,-\overline p_1; \delta,\Lambda)\\&\quad\quad\quad\quad\quad\quad\quad\quad\quad\quad\quad\quad\times a_{\bar{\ell}}^\dagger(\overline p_4) a_{\bar{k}}^\dagger(\overline p_3) a_k(\overline p_2)a_\ell(\overline p_1)\,,\\
 \widehat{\cal H}_4^\text{A-A}&=\int \prod_{i=1}^4 \frac{d^2\overline p_i}{(2\pi)^2} \Theta(p_i^+)(2\pi)^2 \delta^2(\overline p_{14}+\overline p_{23})\frac{1}{2}\cH_4^{\rm reg}(-\overline p_4,\overline p_1,-\overline p_2,\overline p_3; \delta,\Lambda)\\&\quad\quad\quad\quad\quad\quad\quad\quad\quad\quad\quad\quad\times b_{{\ell}}^\dagger(\overline p_4) b_{{k}}^\dagger(\overline p_3) b_{\bar{k}}(\overline p_2)b_{\bar{\ell}}(\overline p_1)\,,\\
\fe where $\cH_4^{\rm reg}$ was written in (\ref{eq:H4} -- \ref{c4counter}). The particle number nonconserving interactions are
\ie
 \widehat{\cal H}_4^{3\to1} &= \int \prod_{i=1}^4 \frac{d^2\overline p_i}{(2\pi)^2} \Theta(p_i^+)(2\pi)^2 \delta^2(\overline p_1+\overline p_2 + \overline p_3-\overline p_4)\,\\&\quad\quad\quad\times \Big(\cH_4^{\rm reg}(\overline p_1,\overline p_2,\overline p_4,-\overline p_3; \delta,\Lambda) \, a^\dagger_{\bar{\ell}}(\overline p_4) a_{\ell}(\overline p_3) b_{\bar{k}}(\overline p_2) a_k(\overline p_1) \\&\quad\quad\quad\quad\quad\quad\quad\quad+\cH_4^{\rm reg}(\overline p_2,\overline p_1,-\overline p_3,\overline p_4; \delta,\Lambda) \, b^\dagger_\ell(\overline p_4) b_{\bar{\ell}}(p_3) a_k(\overline p_2)b_{\bar{k}}(\overline p_1)\Big)\,,\\
\widehat{\cal H}_4^{1\to3} &= \int \prod_{i=1}^4 \frac{d^2\overline p_i}{(2\pi)^2} \Theta(p_i^+)(2\pi)^2 \delta^2( \overline p_4 -\overline p_3-\overline p_2-\overline p_1 )\,\\&\quad\quad\quad\times\Big(\cH_4^{\rm reg}(-\overline p_2,-\overline p_1,\overline p_3,-\overline p_4; \delta,\Lambda) \,a^\dagger_{\bar{\ell}}(\overline p_1) b^\dagger_\ell(\overline p_2) a^\dagger_{\bar{k}}(\overline p_3) a_k(\overline p_4) \\&\quad\quad\quad\quad\quad\quad\quad\quad+\cH_4^{\rm reg}(-\overline p_1,-\overline p_2,-\overline p_4,\overline p_3; \delta,\Lambda)\, b^\dagger_\ell(\overline p_1) a^\dagger_{\bar \ell}(\overline p_2) b^\dagger_{k}(\overline p_3 ) b_{\bar{k}}(\overline p_4)\Big)\,.
\fe

\section{The $T$-matrix from Lippmann-Schwinger equation} \label{app:tmatrix}

Starting with the Lippmann-Schwinger equation for the in/out-states in terms of the free-particle basis states, either (\ref{inoutrelation}) or equivalently
\ie\label{inoutrelationapp}
|\alpha\rangle^\mathrm{in/out} 
&= \Big(1 + \frac{1}{E_{\alpha} - \widehat{\mathcal{H}}\pm i\epsilon}\widehat{V}\Big) |\alpha\rangle^0 ,
\fe
we can rewrite the S-matrix elements as
\ie
{}^\mathrm{out}\langle \beta | \alpha\rangle^\mathrm{in} &= {}^0\langle \beta | \A\rangle^{\rm in} +  {}^0\langle\B| \widehat{V}\frac{1}{E_{\beta} - \widehat{\mathcal{H}} + i\epsilon } | \alpha\rangle^{\rm in} 
\\
&= \,^0\langle \beta | \alpha\rangle^0 +\,^0\langle \beta | \frac{1}{E_{\alpha} - \widehat{\mathcal{H}}_2 + i\epsilon }\widehat{V}|\alpha\rangle^\mathrm{in}  + \frac{1}{E_\beta - E_\alpha + i\epsilon} \,^0\langle \beta|\widehat{V}|\alpha\rangle^\mathrm{in} 
\\
&= \,^0\langle \beta | \alpha\rangle^0 - 2\pi i \delta(E_\A - E_\B) \, {}^0\langle \beta|\widehat{V}|\alpha\rangle^\textrm{in},
\fe
where we used (\ref{inoutrelationapp}) for ${}^{\rm out}\langle\B|$ in the first equality, and (\ref{inoutrelation}) for $|\A\rangle^{\rm in}$ in the second equality. This gives (\ref{smatrixeqn}) with the expression (\ref{tmatrixexpr}) for the $T$-matrix element.

\section{Glossary of notations}
\label{app:relabelled}

In this Appendix we recap various symbols introduced in this paper, and list some additional useful relations. 

In the $2\to 2$ scattering process, we label the incoming momenta by $p_1, p_2$, and the outgoing momenta by $p_3, p_4$. The gauge-singlet particle/anti-particle in-state $|\vec p_1, \vec p_2\rangle^{\rm in}$ is normalized covariantly according to ${}^{\rm in}\langle \vec p_3,\vec p_4 |\vec p_1, \vec p_2\rangle^{\rm in}=I(p_3,p_4|p_1,p_2)$ (\ref{identityi}), and similarly for the out-state $|\vec p_3,\vec p_4\rangle^{\rm out}$.

In lightcone coordinates, the mass-shell relation reads $-p_i^- = {(p_i^\perp)^2+m^2\over 2p_i^+}$.
Working at fixed total lightcone momentum $P^+ = p_1^+ + p_2^+ = p_3^+ + p_4^+$, we parameterize the incoming momenta with $(x,p)$, and the outgoing momentum by $(y,q)$, defined in (\ref{momentasplit}). Further working at a fixed Mandelstam variables $s= -(p_1+p_2)^2 = -(p_3+p_4)^2$, we have labeled the in-states with $(x,a=\pm)$, and the out-states with $(y,b=\pm)$. The transverse momenta $p$ and $q$ are related by
\ie
p = a \, \ell_*(x),~~~~ q = b \, \ell_*(y),
\fe
with $\ell_*$ defined in (\ref{ellresidue}).

The asymptotic states in the lightfront quantization are deonted $|x,p\rangle^{\rm in}$ and $|y,q\rangle^{\rm out}$ respectively. They are related to the covariant basis states by a normalization factor ${\cal N}(\A)$, as in (\ref{xpphase}) and (\ref{yqphase}). The phase of ${\cal N}(\A)$, denoted $e^{i\varphi(\A)}$, is determined through (\ref{varphipha}), (\ref{anyonspin}), and (\ref{eq:0thorderphase}). In other words, we found $\varphi(\A) = (1+\widetilde\lambda)\varphi^{(0)}(\A)$, where $\varphi^{(0)}(\A)$ is the tree-level phase correction (\ref{eq:0thorderphase}).

Alternatively, we can label the in- and out-states in the center-of-mass frame by the angles $\A$ and $\B$, defined in (\ref{comassignment}). They are related to $(x,p)$ and $(y,q)$ by
\ie\label{eq:xtoangle}
&\cos\alpha = {\sqrt{s}\over \sqrt{s-4m^2}} (1-2x),~~~~ {\rm sign}(\sin\alpha)= -{\rm sign}(p),
\\
&\cos\B = {\sqrt{s}\over \sqrt{s-4m^2}} (1-2y),~~~~ {\rm sign}(\sin\B)= -{\rm sign}(q).
\fe
The scattering angle $\theta\equiv \B-\A$ can be expressed in terms of $(x,a)$ and $(y,b)$ through
\ie
\cot {\theta \over 2} &={1\over \sqrt{s}} \left[ -2im +i { x(1-x) y(1-y) \over (x-y) (s-4 m^2)} u_a^-(x) u_b^+(y) (u_a^+(x) - u_b^-(y))\right],
\fe
where $u_a^\pm(x)$ are defined in (\ref{uadef}).

The part of the lightcone Hamiltonian relevant for the planar singlet channel amplitude is (\ref{h24sing}). The interaction $\widehat\cH_4^{\rm S}$ is expanded in terms of the fermion creation and annihilation operators with  coefficient $\cH_4^{\rm reg}$ defined in (\ref{interaction}). The latter is further separated (\ref{eq:H4}) into a classical term $\cH_4^{\rm cl}$ (\ref{H4explicit}) and a counter term ${\cal C}_4$ (\ref{c4counter}). The interaction ``potential" $\cV^{ba}(y|x)$ appearing in the manifestly finite integral equation (\ref{bestintegraleqn}) is none other than a rewriting of $\cH_4^{\rm cl}$ (\ref{vbaexpr}) in terms of the variables $(x,a)$ and $(y,b)$ at fixed $s$.

The $2\to 2$ scattering amplitude in the singlet channel is expressed through the $T$-matrix. The covariant $T$-matrix element $T_{\rm S}(s,\theta)$ is defined in (\ref{sanygen}), and is decomposed in terms of the forward-distribution coefficient $T_0(s)$, and the function part $\widetilde T(s,\theta)$, defined in (\ref{ttapp}). The analogous $T$-matrix elements in the lightfront basis is denoted $\cT(y,q|x,p)$ or equivalently at fixed $s$, $\cT^{ba}(y|x)$, and decomposed into the forward-distribution coefficient $\cT_0(x)$ and the function part $\widetilde \cT^{ba}(y|x)$ via (\ref{eq:genform}). Their relation to the covariant $T$-matrix element is given in (\ref{ttmatch}). The $T$-matrix element at $L$-loop order is denoted, after stripped off the factor $\lambda^{L+1}$, by the same symbol with a superscript $(L)$.


\section{Derivation of the finite integral equation \label{app:finite}}

We begin with the observation that $\delta_C(q,y|p,x;\delta,\Lambda)$ defined in (\ref{c4cancel}) vanishes in the $\delta\to 0$ limit, and is regular at $x=0,1$. Consider the $L$-loop amplitude ($L>1$)
\ie
\mathcal{T}^{(L)}(y, q|x,p) &= \lim_{\Lambda \to \infty} \lim_{\delta \to 0} {(P^+)^2 \over (2\pi)^2} \dashint_{\frac{\delta}{P^+}}^{1-\frac{\delta}{P^+}} dz \int_{-\Lambda}^{\Lambda} d\ell \, G(z, \ell |x,p;\eps)
\\
&~~~ \times \Bigg[\cT^{(0)}(y,q |z, \ell)\mathcal{T}^{(L-1)}(z, \ell | x, p)
	+ \lambda^{-2} \Vc(y,q|z. \ell ;\delta,\Lambda)\mathcal{T}^{(L-2)}(z, \ell | x, p) \Bigg],
\fe 
where 
\beq
G(z, \ell|x, p;\epsilon)=\bigg[ \frac{p^2+m^2}{2 x (1-x)} - \frac{\ell^2+m^2}{2 z (1-z)} + i\epsilon \bigg]^{-1}  .
\eeq
Using (\ref{c4cancel}), 
we can write
\ie\label{tltep}
\mathcal{T}^{(L)}(y,q|x,p) &= \lim_{\Lambda \to \infty} \lim_{\delta \to 0}  {(P^+)^2\over (2\pi)^2} \dashint_{\frac{\delta}{P^+}}^{1-\frac{\delta}{P^+}} dz  \int_{-\Lambda}^{\Lambda} d\ell \,G(z, \ell |x,p;\epsilon)\Bigg[ \cT^{(0}(y,q|z,\ell)\mathcal{T}^{(L-1)}(z,\ell|x, p)
\\
&~~~~~ + {\Lambda (P^+)^2\over 2\pi^2} \dashint_{\frac{\delta}{P^+}}^{1-\frac{\delta}{P^+}} dz' \,\,  {2z'(1-z') \overline\cT^{(0)}_\infty(y|z') \cT^{(0)}_\infty(z'|z)} \, \mathcal{T}^{(L-2)}(z,\ell|x,p)
\\
&~~~~~+ \lambda^{-2} \delta_C(y,q|z,\ell;\delta,\Lambda)\mathcal{T}^{(L-2)}(z,\ell|x,p) \Bigg] .
\fe
Assuming that $\mathcal{T}$ has at most a simple pole at $z= 0,1$, which will be justified a posteriori, together with the facts that $K$ vanishes at $z=0,1$ and that $\delta_C$ is regular in the integration range, we conclude that the last term in the bracket of (\ref{tltep}) has a vanishing contribution. Next, we change the order of integration in the second line of (\ref{tltep}) and write
\ie{}
&\mathcal{T}^{(L)}(y,q|x,p) = \lim_{\Lambda \to \infty}\lim_{\delta \to 0} \Bigg[ {(P^+)^2\over (2\pi)^2} \dashint_{\frac{\delta}{P^+}}^{1-\frac{\delta}{P^+}} dz \int_{-\Lambda}^{\Lambda} d\ell \,G(z,\ell|x,p;\epsilon) \cT^{(0)}(y,q|z,\ell) \mathcal{T}^{(L-1)}(z,\ell|x,p)
\\
&+ {\Lambda (P^+)^4\over 8\pi^4}\,\dashint_{
		\frac{\delta}{P^+}}^{1-\frac{\delta}{P^+}} dz' \,\,  2z'(1-z') \overline\cT^{(0)}_\infty(y|z') \dashint_{
		\frac{\delta}{P^+}}^{1-\frac{\delta}{P^+}} dz \int_{-\Lambda}^{\Lambda} d\ell \,G(z,\ell|x,p;\epsilon) \cT^{(0)}_\infty(z'|z) \mathcal{T}^{(L-2)}(z,\ell|x,p) \Bigg].
\fe
In the second line, the $z'$-integral can at most produce a log divergence in $\delta$, while the $z$-integral gives a non-singular power series in $\delta$. Furthermore, the $\ell$-integral gives a result that is finite in the $\Lambda\to\infty$ limit. Therefore, we can take the limit on the $z,\ell$ integral, giving
\ie\label{tloop}
& \mathcal{T}^{(L)}(y,q|x,p) = \lim_{\Lambda \to \infty}\lim_{\delta \to 0} \Bigg[ {(P^+)^2 \over (2\pi)^2} \dashint_{\frac{\delta}{P^+}}^{1-\frac{\delta}{P^+}} dz \int_{-\Lambda}^{\Lambda} d\ell \,G(z,\ell|x,p;\epsilon) \cT^{(0)}(y,q|z,\ell) \mathcal{T}^{(L-1)}(z,\ell|x,p)
\\
& + {\Lambda (P^+)^4\over 8\pi^4} \dashint_{\frac{\delta}{P^+}}^{1-\frac{\delta}{P^+}} dz' \,  2z'(1-z') \overline\cT^{(0)}_\infty(y|z') \dashint_{0}^{1} dz \int_{-\infty}^{\infty} d\ell \,G(\ell,z|x,p;\epsilon) \cT^{(0)}_\infty(z'|z) \mathcal{T}^{(n-2)}(z,\ell|x,p) \Bigg] .
\fe 
Next, we reintroduce a trivial $\ell'$-integral, and then relabel $z',\ell' \leftrightarrow z, \ell$,
\ie\label{eq:finEqnloop}
& \mathcal{T}^{(L)}(y,q|x,p) = \lim_{\Lambda \to \infty} \lim_{\delta \to 0} {(P^+)^2\over (2\pi)^2} \dashint_{\frac{\delta}{P^+}}^{1-\frac{\delta}{P^+}} dz \int_{-\Lambda}^{\Lambda} d\ell \,G(z,\ell|x,p;\epsilon) \Bigg[ \cT^{(0)}(y,q|z,\ell) \mathcal{T}^{(L-1)}(z,\ell|x,p)
\\
& ~~~ + {(P^+)^2\over 2\pi^2} z(1-z)\overline\cT^{(0)}_\infty(y|z) \dashint_{0}^{1} dz' \int_{-\infty}^{\infty} d\ell' \, G(z', \ell' | x,p;\epsilon) \, \cT^{(0)}_\infty(z|z') \mathcal{T}^{(L-2)}(z',\ell'|x,p)\Bigg].
\fe 
By inspecting (\ref{eq:intEqT1_unregularized}) and the definition (\ref{eq:Tinfty}), we have
\ie
\mathcal{T}_\infty(z|x,p) &= \lim_{\Lambda \to \infty} \lim_{\delta \to 0} \Bigg[ \lambda \cT^{(0)}_\infty(z|x)+ \lambda
{(P^+)^2\over (2\pi)^2} \dashint_{\frac{\delta}{P^+}}^{1-\frac{\delta}{P^+}} dz' \int_{-\Lambda}^{\Lambda} d\ell' \,G(\ell',z';\eps) \cT^{(0)}_\infty(z|z') \mathcal{T}(z',\ell'|x,p) \Bigg],
\fe
We see that (\ref{tloop}) is precisely $L$-loop (i.e. order $\lambda^{L+1}$) part of (\ref{eq:intEqT1}), for $L>1$.


\section{Numerical check of the all-order solution}
\label{allorderstest}

In this appendix, we verify numerically that the analytic ansatz (\ref{ansatztilde}), together with the coupling redefinition (\ref{couplingredef}), satisfies the integral equation (\ref{secondbestintegraleqn}). This is performed by a direct comparison of the two sides of (\ref{secondbestintegraleqn}) evaluated with the ansatz.


\begin{figure}[h!]
	\def\svgwidth{1\linewidth}
	\centering{
		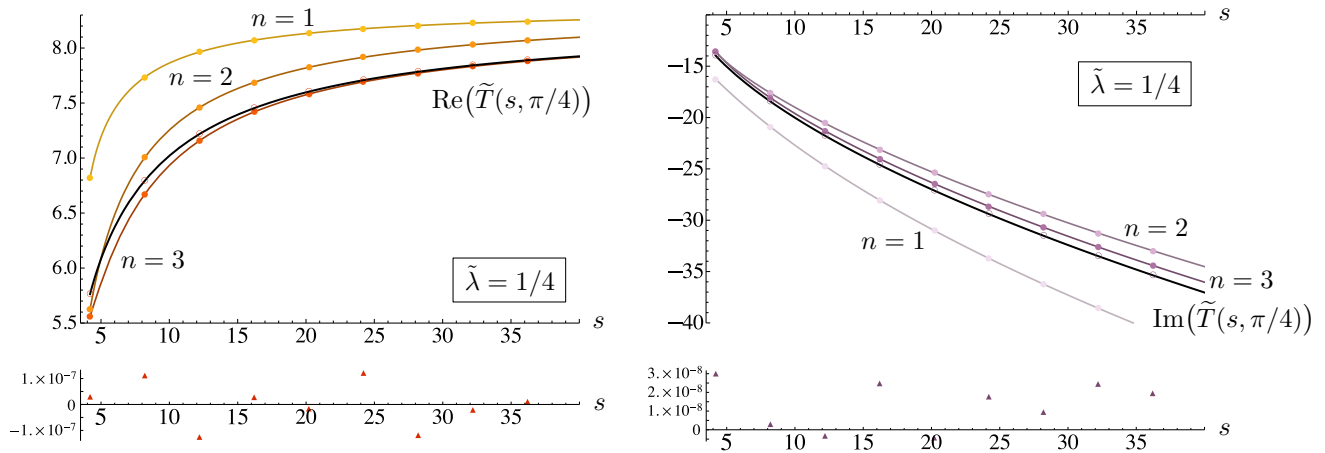	\caption{Numerical tests of the integral equation at coupling $\tilde{\lambda}=1/4$ and scattering angle $\theta = \beta-\alpha = \pi/4$ for a range of $s$ (having set $m=1$). The left panel compares the real part of $\widetilde{T}(s,\theta)$ (black) to $\sum_{L=0}^n \lambda^{L+1} \widetilde T^{(L)}(s,\theta)$ (orange) up to three-loop order. The right panel compares the imaginary part, with all-order results in black, and perturbative results in purple. The deviation between the two sides of the integral equation for the all-orders ansatz is shown in the lower panels.
			\label{weakcoupling}
	}}
\end{figure}

Figure \ref{weakcoupling} shows such a numerical test for the real and imaginary part of the function part of the amplitude in the covariant basis, $\widetilde T(s,\theta)$, at $\widetilde\lambda={1\over 4}$ and scattering angle $\theta={\pi\over 4}$ over a range of $s$. As Lorentz invariance is not manifest in (\ref{secondbestintegraleqn}), we have chosen the incoming angle $\A={\pi\over 2}$ and outgoing angle $\B={3\pi\over 4}$. The numerical integral is evaluated using the ``PrincipalValue" method of NIntegrate in Mathematica. The agreement with the integral equation is well within 7 significant digits. We have also included a comparison with the perturbative results up to $n$-loop ordfer, for $n=1,2,3$.  The perturbative convergence supports the claim of a unique solution to (\ref{bestintegraleqn}).

\begin{figure}[h!]
	\def\svgwidth{1\linewidth}
	\centering{
		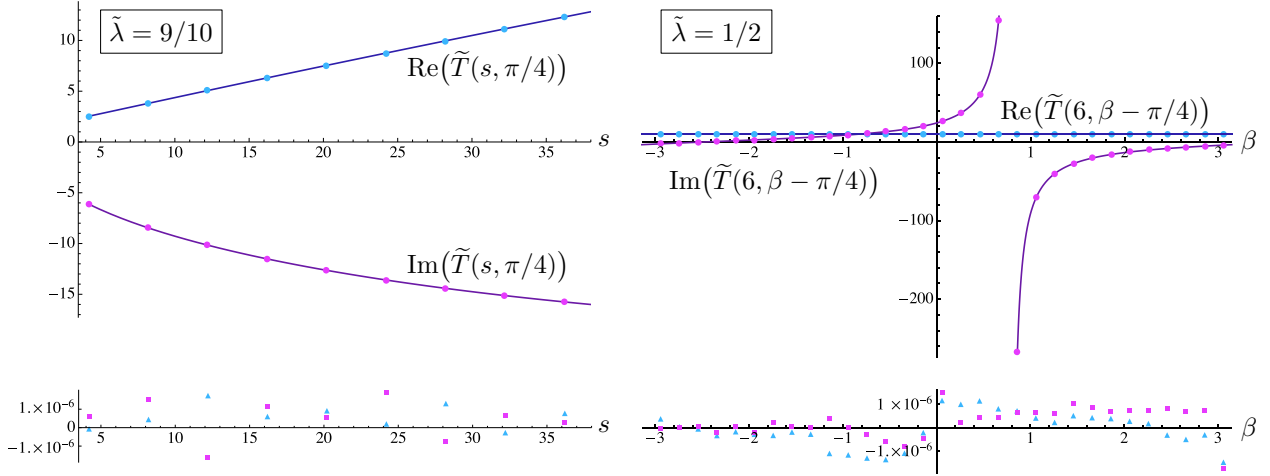	\caption{Additional numerical tests comparing two sides of the integral equation (\ref{bestintegraleqn}) with the analytic ansatz. The left panel shows strong coupling data in the same kinematic setup as that of Figure \ref{weakcoupling}. The right panel shows data at fixed center-of-mass energy, with $s= 6$, and fixed incoming angle $\alpha=\pi/4$. The deviation between two sides of the integral equation, for the real and imaginary parts of $\widetilde T$, are shown in the lower panels.
			\label{strongcoupling}
	}}
\end{figure}

Figure \ref{strongcoupling} shows an analogous plot at strong coupling $\widetilde\lambda={9\over 10}$, and a plot of the angular dependence at $\widetilde\lambda={1\over 2}$. The agreement is within 6 significant digits.

\bibliography{refs}

\providecommand{\href}[2]{#2}\begingroup\raggedright\begin{thebibliography}{10}

\bibitem{Deser:1981wh}
S.~Deser, R.~Jackiw, and S.~Templeton, {\it {Topologically Massive Gauge
  Theories}},  {\em Annals Phys.} {\bf 140} (1982) 372--411. [Erratum: Annals
  Phys. 185, 406 (1988)].

\bibitem{Deser:1982vy}
S.~Deser, R.~Jackiw, and S.~Templeton, {\it {Three-Dimensional Massive Gauge
  Theories}},  {\em Phys. Rev. Lett.} {\bf 48} (1982) 975--978.

\bibitem{Witten:1988hf}
E.~Witten, {\it {Quantum Field Theory and the Jones Polynomial}},  {\em Commun.
  Math. Phys.} {\bf 121} (1989) 351--399.

\bibitem{Fredenhagen:1988fj}
K.~Fredenhagen, K.-H. Rehren, and B.~Schroer, {\it {Superselection Sectors with
  Braid Group Statistics and Exchange Algebras. 1. General Theory}},  {\em
  Commun. Math. Phys.} {\bf 125} (1989) 201.

\bibitem{Frohlich:1990ww}
J.~Frohlich and F.~Gabbiani, {\it {Braid statistics in local quantum theory}},
  {\em Rev. Math. Phys.} {\bf 2} (1991) 251--354.

\bibitem{Frohlich:1990xz}
J.~Frohlich and T.~Kerler, {\it {Universality in quantum Hall systems}},  {\em
  Nucl. Phys. B} {\bf 354} (1991) 369--417.

\bibitem{Frohlich:1991wb}
J.~Frohlich and A.~Zee, {\it {Large scale physics of the quantum Hall fluid}},
  {\em Nucl. Phys. B} {\bf 364} (1991) 517--540.

\bibitem{Iengo:1991zbc}
R.~Iengo and K.~Lechner, {\it {Anyon quantum mechanics and Chern-Simons
  theory}},  {\em Phys. Rept.} {\bf 213} (1992) 179--269.

\bibitem{Kitaev:2005hzj}
A.~Kitaev, {\it {Anyons in an exactly solved model and beyond}},  {\em Annals
  Phys.} {\bf 321} (2006), no.~1 2--111,
  [\href{http://arxiv.org/abs/cond-mat/0506438}{{\tt cond-mat/0506438}}].

\bibitem{Aharony:2011jz}
O.~Aharony, G.~Gur-Ari, and R.~Yacoby, {\it {d=3 Bosonic Vector Models Coupled
  to Chern-Simons Gauge Theories}},  {\em JHEP} {\bf 03} (2012) 037,
  [\href{http://arxiv.org/abs/1110.4382}{{\tt arXiv:1110.4382}}].

\bibitem{Giombi:2011kc}
S.~Giombi, S.~Minwalla, S.~Prakash, S.~P. Trivedi, S.~R. Wadia, and X.~Yin,
  {\it {Chern-Simons Theory with Vector Fermion Matter}},  {\em Eur. Phys. J.
  C} {\bf 72} (2012) 2112, [\href{http://arxiv.org/abs/1110.4386}{{\tt
  arXiv:1110.4386}}].

\bibitem{Aharony:2012ns}
O.~Aharony, S.~Giombi, G.~Gur-Ari, J.~Maldacena, and R.~Yacoby, {\it {The
  Thermal Free Energy in Large N Chern-Simons-Matter Theories}},  {\em JHEP}
  {\bf 03} (2013) 121, [\href{http://arxiv.org/abs/1211.4843}{{\tt
  arXiv:1211.4843}}].

\bibitem{Jain:2013gza}
S.~Jain, S.~Minwalla, and S.~Yokoyama, {\it {Chern Simons duality with a
  fundamental boson and fermion}},  {\em JHEP} {\bf 11} (2013) 037,
  [\href{http://arxiv.org/abs/1305.7235}{{\tt arXiv:1305.7235}}].

\bibitem{Jain:2014nza}
S.~Jain, M.~Mandlik, S.~Minwalla, T.~Takimi, S.~R. Wadia, and S.~Yokoyama, {\it
  {Unitarity, Crossing Symmetry and Duality of the S-matrix in large N
  Chern-Simons theories with fundamental matter}},  {\em JHEP} {\bf 04} (2015)
  129, [\href{http://arxiv.org/abs/1404.6373}{{\tt arXiv:1404.6373}}].

\bibitem{Aharony:2015mjs}
O.~Aharony, {\it {Baryons, monopoles and dualities in Chern-Simons-matter
  theories}},  {\em JHEP} {\bf 02} (2016) 093,
  [\href{http://arxiv.org/abs/1512.00161}{{\tt arXiv:1512.00161}}].

\bibitem{Kruczenski:2022lot}
M.~Kruczenski, J.~Penedones, and B.~C. van Rees, {\it {Snowmass White Paper:
  S-matrix Bootstrap}},  \href{http://arxiv.org/abs/2203.02421}{{\tt
  arXiv:2203.02421}}.

\bibitem{tHooft:1974pnl}
G.~'t~Hooft, {\it {A Two-Dimensional Model for Mesons}},  {\em Nucl. Phys. B}
  {\bf 75} (1974) 461--470.

\bibitem{Dalley:1992yy}
S.~Dalley and I.~R. Klebanov, {\it {String spectrum of (1+1)-dimensional large
  N QCD with adjoint matter}},  {\em Phys. Rev. D} {\bf 47} (1993) 2517--2527,
  [\href{http://arxiv.org/abs/hep-th/9209049}{{\tt hep-th/9209049}}].

\bibitem{Bhanot:1993xp}
G.~Bhanot, K.~Demeterfi, and I.~R. Klebanov, {\it {(1+1)-dimensional large N
  QCD coupled to adjoint fermions}},  {\em Phys. Rev. D} {\bf 48} (1993)
  4980--4990, [\href{http://arxiv.org/abs/hep-th/9307111}{{\tt
  hep-th/9307111}}].

\bibitem{Dempsey:2021xpf}
R.~Dempsey, I.~R. Klebanov, and S.~S. Pufu, {\it {Exact symmetries and
  threshold states in two-dimensional models for QCD}},  {\em JHEP} {\bf 10}
  (2021) 096, [\href{http://arxiv.org/abs/2101.05432}{{\tt arXiv:2101.05432}}].

\bibitem{Delacretaz:2018xbn}
L.~V. Delacr\'etaz, A.~L. Fitzpatrick, E.~Katz, and L.~G. Vitale, {\it
  {Conformal Truncation of Chern-Simons Theory at Large $N_f$}},  {\em JHEP}
  {\bf 03} (2019) 107, [\href{http://arxiv.org/abs/1811.10612}{{\tt
  arXiv:1811.10612}}].

\bibitem{Weinberg:1995mt}
S.~Weinberg, {\em {The Quantum theory of fields. Vol. 1: Foundations}}.
\newblock Cambridge University Press, 6, 2005.

\bibitem{FROHLICH1991533}
J.~Fröhlich and P.~Marchetti, {\it Spin-statistics theorem and scattering in
  planar quantum field theories with braid statistics},  {\em Nuclear Physics
  B} {\bf 356} (1991), no.~3 533--573.

\bibitem{Anand:2020gnn}
N.~Anand, A.~L. Fitzpatrick, E.~Katz, Z.~U. Khandker, M.~T. Walters, and
  Y.~Xin, {\it {Introduction to Lightcone Conformal Truncation: QFT Dynamics
  from CFT Data}},  \href{http://arxiv.org/abs/2005.13544}{{\tt
  arXiv:2005.13544}}.

\bibitem{Fitzpatrick:2022dwq}
A.~L. Fitzpatrick and E.~Katz, {\it {Snowmass White Paper: Hamiltonian
  Truncation}},  \href{http://arxiv.org/abs/2201.11696}{{\tt
  arXiv:2201.11696}}.

\end{thebibliography}\endgroup


\providecommand{\href}[2]{#2}\begingroup\raggedright\endgroup
\bibliographystyle{JHEP}
\end{document}